\documentclass[graybox,natbib]{svmult}
\usepackage[utf8x]{inputenc} % funny chars supported
\usepackage{newtxtext} % Should be used according to Elsevier
\usepackage{newtxmath} % Should be used according to Elsevier
\usepackage[bottom]{footmisc} % Should be used according to Elsevier

\usepackage{type1cm}        % activate if the above 3 fonts are
\usepackage{makeidx}         % allows index generation
\usepackage{graphicx}        % standard LaTeX graphics tool
\usepackage{url}

\usepackage{framed,color}
\definecolor{shadecolor}{rgb}{.5,.5,.5} % change to suitable shade of grey

\DeclareMathOperator{\logit}{logit}

\newcommand{\bigCI}{\mathrel{\text{\scalebox{1.07}{$\perp\mkern-10mu\perp$}}}}

\usepackage{tabularx} % Easier to adjust tables
\usepackage{listings} % code listings [not sure we need it]
\usepackage{IEEEtrantools} % For lsiting math equations and aligning them

\usepackage[caption=false,font=footnotesize]{subfig}

\usepackage{siunitx}
\usepackage{upquote}
\usepackage{listings}

\newcommand{\ESE}{empirical software engineering}

\makeindex             % used for the subject index
\begin{document}

\title*{Bayesian data analysis in empirical software engineering---The case of missing data}
\author{Richard Torkar, Robert Feldt, and Carlo A.\ Furia}
\institute{Richard Torkar and Robert Feldt \at Chalmers and University of Gothenburg, SE-412 96 Gothenburg, Sweden \email{torkarr@chalmers.se, robert.feldt@chalmers.se}
\and Carlo A.\ Furia \at Universit\`a della
Svizzera italiana,
Via Giuseppe Buffi 13,
I-6904 Lugano,
Switzerland \email{furiac@usi.ch}}

\maketitle

\abstract*{Bayesian data analysis (BDA) is today used by a multitude of research disciplines. These disciplines use BDA as a way to embrace uncertainty by using multilevel models and making use of all available information at hand. In this chapter, we first introduce the reader to BDA and then provide an example from empirical software engineering, where we also deal with a common issue in our field, i.e., missing data. 
The example we make use of presents the steps done when conducting state of the art statistical analysis. First, we need to understand the problem we want to solve. Second, we conduct causal analysis. Third, we analyze non-identifiability. Fourth, we conduct missing data analysis. Finally, we do a sensitivity analysis of priors. All this before we design our statistical model. Once we have a model, we present several diagnostics one can use to conduct sanity checks.
We hope that through these examples, the reader will see the advantages of using BDA\@. This way, we hope Bayesian statistics will become more prevalent in our field, thus partly avoiding the reproducibility crisis we have seen in other disciplines.}

\abstract{Bayesian data analysis (BDA) is today used by a multitude of research disciplines. These disciplines use BDA as a way to embrace uncertainty by using multilevel models and making use of all available information at hand. In this chapter, we first introduce the reader to BDA and then provide an example from empirical software engineering, where we also deal with a common issue in our field, i.e., missing data. 
The example we make use of presents the steps done when conducting state of the art statistical analysis. First, we need to understand the problem we want to solve. Second, we conduct causal analysis. Third, we analyze non-identifiability. Fourth, we conduct missing data analysis. Finally, we do a sensitivity analysis of priors. All this before we design our statistical model. Once we have a model, we present several diagnostics one can use to conduct sanity checks.
We hope that through these examples, the reader will see the advantages of using BDA\@. This way, we hope Bayesian statistics will become more prevalent in our field, thus partly avoiding the reproducibility crisis we have seen in other disciplines.}

\section{Introduction}
\label{sec:intro}
Statistics, we argue, is one of the principal tools researchers in empirical software engineering have at their disposal to build an argument that guides them towards the ultimate objective, i.e., practical significance and (subsequent) impact of their findings\footnote{In this chapter, we focus on \ESE{} research where quantitative data is a major component; for studies that are mainly qualitative a different set of concerns need to be taken into account, see for example~\citep{lenberg2017bse_qualitative}}. Practical significance is, as we have seen~\citep{TorkarFNG17arxiv}, not very often explicitly discussed in software engineering research today and we argue that this is mainly out of two reasons.

The first one being that statistical maturity of \ESE{} research is not high enough~\citep{TorkarFNG17arxiv}, leading to difficulties with connecting statistical findings to practical significance. The second reason is a combination of issues hampering our research field, e.g., small sample sizes, failure to analyze disparate types of data in a unified framework or lack of data availability (only 13\% of publications provide a replication package and carefully describe each step to make reproduction feasible~\citep{rodriguez18repro}).

Both of the above issues are worrisome since it could make it hard to strengthen arguments concerning practical significance, e.g., connecting effort and, conclusively, ROI\footnote{In literature, Return-On-Investment refers to, in various ways, the calculation one does to see the benefit (\emph{return}) an investment (\emph{cost}) has.} to the findings of a research study, if one would want so. For academic research to be more relevant and have more impact on practitioners, its practical significance and its implications need to be precise.

Furthermore, issues such as the above is also likely to lead \ESE{} towards a replication crisis as we have seen in other disciplines, e.g., medicine~\citep{ioannidis05false,ioannidis05contra,ioannidis16false, Glick92account}, psychology~\citep{aa15crisis,john12qrp,shanks13priming}, economics~\citep{ioannidsSD17bias, Camerer16repl}, and marketing~\citep{hunter01repl}.

In order to solve some of the above challenges researchers have proposed that we need to focus on, e.g., (i) openness, i.e., that data and manuscripts are accessible for all stakeholders, (ii) preregistration, i.e., a planned study is peer-reviewed in the usual manner and accepted by a journal \textit{before} the experiment is run, so that there is no incentive to look for significance after-the-fact~\citep{dutilh2017prereg}, (iii) increasing the sample size, (iv) lowering the significance threshold from $p<0.05$ to $p<0.005$~\citep{benjamin17ssd}, and (v) removing null hypothesis significance testing (NHST) altogether, which the journal \textit{Basic and Applied Social Psychology} advocates~\citep{TrafimowM15}, as do~\citet{gelman17abandon}.

However, some researchers, most notably~\citet{gelman18failure}, claim that even the above is not enough and argue that a unified approach for these matters should mainly evolve from three components: Procedural solutions, solutions based on design and data collection, and improved statistical analysis.

Concerning procedural solutions,~\citet{gelman18failure} like others, suggests publishing papers on, e.g., Arxiv, to encourage post-publication review, and to use preregistration as a tool for lowering the `file drawer' bias. For design and data collection, Gelman provides convincing arguments that we should focus on reducing measurement error (the example being that reducing the measurement error by a factor of two is like multiplying the sample size by a factor of four), and move to within-subject from between-subject study designs when possible.\footnote{In a within-subject design the same group of subjects are used in more than one treatment.} Finally, concerning improved statistical analysis, Gelman advocates the use of Bayesian inference and multilevel models (MLMs),\footnote{Multilevel models can also be called hierarchical linear models, nested data models, mixed models, random coefficient, random-effects models, random parameter models, or split-plot designs.} as a way to discuss ``\ldots the range of applicability of a study'', i.e., practical significance.

Overall, we side with these arguments and believe they are critical also for software engineering to better connect empirical research with the practice it ultimately aims to improve. We will thus introduce and exemplify the use of Bayesian statistical methods in \ESE{} research. They are a good starting point since individual researchers can learn them and apply them in isolation without waiting for the community as a whole to take further steps needed to avoid a replication crisis and to become more practically relevant. We argue that using Bayesian methods allows us to better connect our findings to practical significance through the use of more balanced out-of-sample predictions, i.e., one of the outputs from Bayesian data analysis (this will be further elaborated on in Sect.~\ref{sec:short-BDA}). Additionally, we have yet to face data from empirical software engineering where Bayesian data analysis can not be employed, and when having a small sample size, due to the priors employed, Bayesian data analysis, we would argue, shows its strengths.

In this chapter, we rely on three key concepts: Bayes' theorem, multilevel models, and Markov chain Monte Carlo sampling.

Bayes' theorem states that,

\begin{equation}
P(A\vert B) = \frac{P(B\vert A) P(A)}{P(B)}
\end{equation}

where $A$ and $B$ are events, and $P(B)\neq 0$. In the theorem, we have two conditional probabilities, $P(A\vert B)$ and $P(B\vert A)$, the likelihood of event $A$ occurring given that $B$ is true, and vice versa. The marginal probability is then observing $A$ and $B$ independently of each other, i.e., $P(A)$ and $P(B)$. Often the above is rewritten as, $P(A\vert B) \varpropto P(B\vert A) \times P(A)$, i.e., the posterior is proportional to the likelihood times the prior or, in other words, given a likelihood and a prior we will be able to approximate the posterior probability distribution; this is, of course, also applicable to MLMs. We will come back to these concepts in the next section.

Multilevel models are not a particularly new thing. However, in the last decades, they have become accessible to researchers due to the rise in computational power, and they go nicely in hand with Bayesian analysis. Bayesian MLMs have several advantages~\citep{mcelreath15statrethink}: (i) When using repeated sampling they do not underfit or overfit the data to the extent single-level models do (i.e., maximally), (ii) the uncertainty across uneven sample sizes is handled automatically, (iii) they model variation explicitly (between and within clusters of data), and (iv) they preserve uncertainty and makes much data transformation unnecessary. In our particular case, Bayes' theorem is the foundation for conducting inference when using MLMs, and Markov chain Monte Carlo (MCMC) is the engine that drives it.

The reason for using MCMC for sampling is simply that before MCMC was introduced, it was virtually impossible to sample large Bayesian multilevel models~\citep{banerjee2014hierarchical}. Today, if one wants to sample from a complex, multidimensional, unknown, posterior probability distribution, MCMC is a widespread technique to use since we have the computational power available. (For more background on sampling algorithms please see~\cite[Ch.\ 8]{mcelreath15statrethink}.)

Next, we first introduce the main elements of Bayesian Data Analysis (BDA) with a non-Software Engineering example. Our main contribution is then a detailed worked case study of applying BDA to an estimation problem in \ESE{}. In particular, the example highlights that with this BDA analysis, we do not need to delete data points for which some data is missing. We conclude the chapter by discussing the methodological implications.

\begin{shaded} Bayesian data analysis (BDA) is growing and, as such, is being used in many disparate scientific disciplines. The approach of BDA that we use in this chapter relies on designing a generative model which we then can use to do out of sample predictions. It will be a more involved analysis, but in the end we hope that it will also provide us with a richer understanding of the phenomena under study.\end{shaded}

\section{A Short Introduction to Bayesian Data Analysis}\label{sec:short-BDA}
Lately, many tools and probabilistic programming languages have been developed to tackle some of the challenges we face when designing more powerful statistical models. In our view, several things have improved. First, probabilistic programming languages, e.g., using \texttt{Julia} with \texttt{Turing.jl}, or \texttt{Stan}, tailored for statistical programming, in combination with resampling techniques, have matured.\footnote{See \url{https://julialang.org}, \url{http://turing.ml}, and \url{https://mc-stan.org}.} Second, resampling techniques based on MCMC have improved~\citep{brooks2011handbook}. Third, procedures for using these techniques now exist~\citep{talts2018arXiv, betancourt2018arXiv, GabrySVBG17, gelman17likelihood, betancourt17hmc} and are being improved iteratively~\citep{vehtariGSCB19diag}. Together, these development make more powerful analysis methods available to a wider audience. 

In this section, we will provide a short introduction to model design, its tool support, and some terminology that we will use in this chapter. To keep it simple and general, we will take data and an example from everyday life, rather than an \ESE{} example. We do not expect the reader to be an expert after this, but rather be able to follow what we present in this chapter, be better prepared for the \ESE{} case study that then follows, and then perhaps read further into the literature we present in Sect.~\ref{sec:reading}. Let us start with terminology.

In this chapter, we will design statistical models. We will use mathematical notation for precision as well as brevity. To generalize, a model will consist of a likelihood, a linear equation, and priors. The purpose of the model is ultimately to make predictions\slash inferences concerning the outcome by using a posterior predictive distribution. Let us introduce a simple example inspired by~\citet{mcelreath15statrethink}.

We want to predict the height of human beings given their weight. A model could then look like this,

{\footnotesize
\begin{IEEEeqnarray}{rCl}
\mathrm{height}_i & \sim & \mathcal{N}(\mu_i, \sigma)\nonumber\\
\mu_i & = & \alpha + \beta_w \times \mathrm{weight}_i\nonumber\\
\alpha & \sim & \mathcal{N}(181, 20)\nonumber\\
\beta_w & \sim & \mathcal{N}(0, 10)\nonumber\\
\sigma & \sim & \text{Half-Cauchy}(0, 10)\nonumber
\end{IEEEeqnarray}
}

Let us now go through this line by line. First, we claim that height has a Normal distribution with mean $\mu$ and standard deviation $\sigma$, i.e., our likelihood. The subset $i$ in height, weight, and $\mu$ is an indication that this holds for each height we have in the data set, i.e., for every human being in the dataset. But why a Normal distribution? Well, there are ontological and epistemological reasons for this~\citep{mcelreath15statrethink}, but in short: if we add together random values from the same distribution it converges to a normal distribution. Since there are many different factors that, jointly, determines the height of a person, e.g., their genetics, nutrients of the mother during pregnancy, food intake as a small child, etc., and their effects `add up', it is often a sensible assumption to assume the result will be normally distributed.

The next line encodes our main assumption about the heights, i.e., they have a linear connection to the weight (our linear equation). We have an intercept labeled $\alpha$, expressing the average height of a human that has average weight, together with a slope $\beta_w$, which captures how much longer (shorter) a human can be expected to be for each added (subtracted) unit of weight they have. We want to estimate these two \textit{parameters} using the data: height and weight. In this example, height is the \textit{outcome} and weight is the \textit{predictor}. We can have more than one outcome, this is known as a multivariate model (compared to univariate models as in the example above), and we can have more than one predictor, as we will see later in this chapter.

\begin{figure}[t]
    \sidecaption[t]
    \includegraphics[scale=0.34]{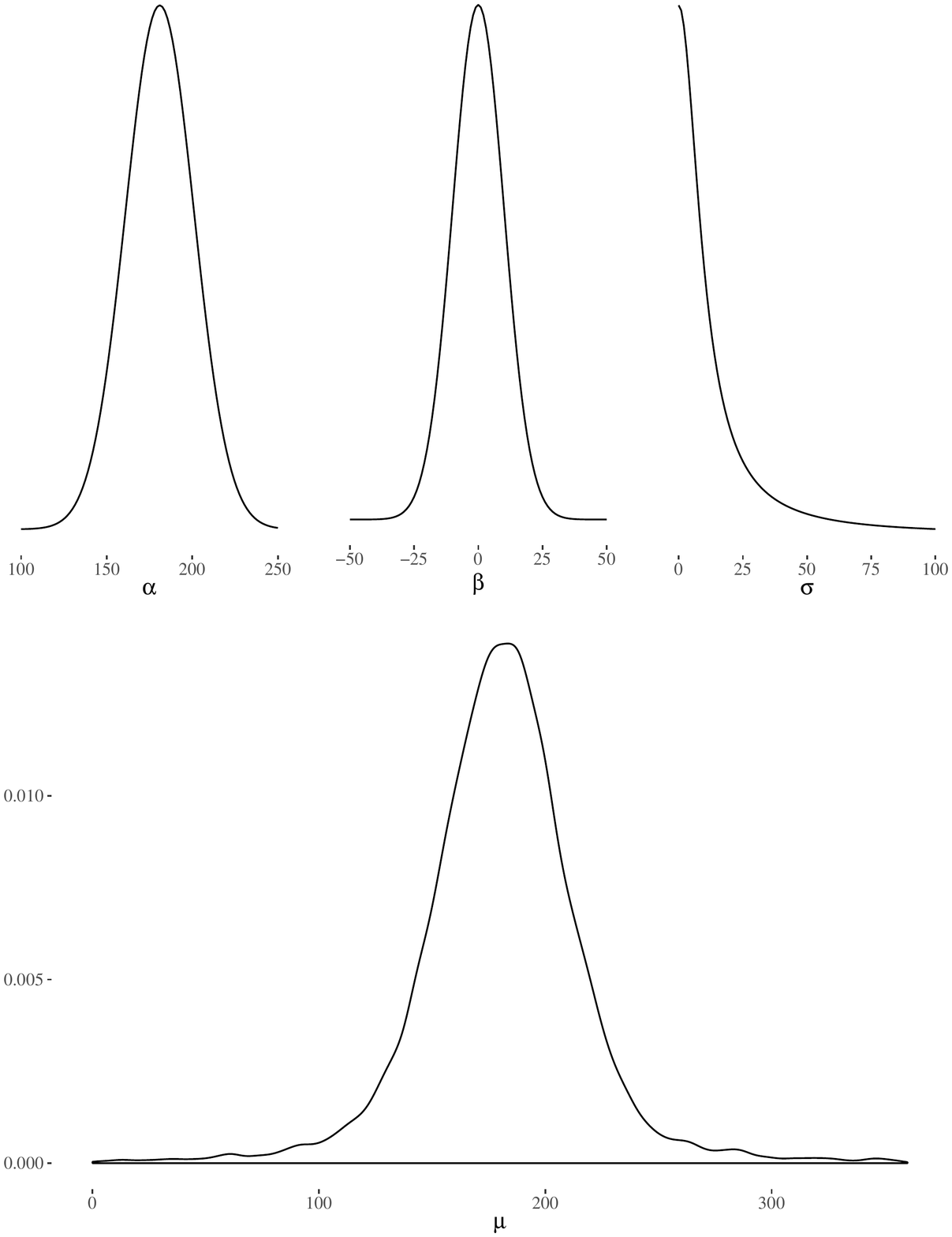}
    \caption{Our selected prior probabilities (priors) for parameters $\alpha$, $\beta$, and $\sigma$, respectively (top row). These are then combined into our prior predictive distribution for the height $\mu$ (bottom row)}
    \label{fig:prior_analysis}
\end{figure}

Next, in a Bayesian model, we need to express our prior belief, our so-called `priors'. The $\alpha$ parameter is the intercept, and hence captures the mean height we expect (see Fig.~\ref{fig:prior_analysis} for a graphical presentation). What we are saying is that we have prior knowledge, i.e., we believe that the mean height will be 181 cm. Why 181? Well, this is the average height of the three authors of this chapter, and when writing the chapter, we had direct and reliable access to this data. Also, our prior expresses that we can expect the mean to vary with a standard deviation of 20, i.e., finding humans with a height in the range 161 to 201 cm would not be too uncommon even if values outside that range can also sometimes happen. For $\beta_w$, our prior indicates that the slope has a mean of 0 and a standard deviation of 10 (Fig.~\ref{fig:prior_analysis}). We could also set a more specific prior here, e.g., we have a feeling that an increase in weight also leads to an increase in height, but let's use a very wide and `allowing' prior.

Finally, we have a prior on $\sigma$, which is the expected variation in actual, measured data from what our linear, `core' model predicts. We have chosen a Half-Cauchy distribution here since we know that the variation cannot be negative (Fig.~\ref{fig:prior_analysis}). The Half-Cauchy is a common prior for $\sigma$ and is roughly a Normal distribution cut in half, i.e., we do not allow negative values on $\sigma$.\footnote{Other priors for $\sigma$ can of course also be used. Please see \url{https://github.com/stan-dev/stan/wiki/Prior-Choice-Recommendations}.} It also has a higher probability for larger values than a Normal distribution, since we have less information on how much variation to expect. In the end, if we have enough evidence (data), it will dominate, also known as `swamp', the priors. This means that the priors are not as critical in situations when we collect lots of data. Before we go on and use our statistical model on actual, measured data, we should study how our model behaves based on only the priors. We can do this by sampling from our priors and `executing' the model to see which heights it predicts. 

In the lower half of Fig.~\ref{fig:prior_analysis}, we see the joint prior probability distribution, which is a combination of the figures on the first row. The mixtures of the priors for $\alpha$, $\beta$, and $\sigma$, and the linear regression they imply could be seen as representative for our height given a weight.

So what does our prior probability distribution (Fig.~\ref{fig:prior_analysis}) tell us? Well, 2.93\% of the population is assumed to be more than 272 cm tall.\footnote{The tallest man, for whom there is irrefutable evidence, was 272 cm.} Additionally, 13.8\% are less than 147 cm.\footnote{People of short stature are \textless147 cm} This seems a bit strange in our view, and even more bizarre is that 0.015\% of the population is shorter than 20 cm, when the shortest human recorded was approximately 53 cm. 

But there is no need to worry. The main idea when selecting priors is to delimit the volume that the sampling needs to cover. We want to get rid of obviously absurd values while ensuring that we do not rule out values that could happen. Who knows, maybe someone who is \textless53 cm or \textgreater272 cm will be found this year. We have just conducted a \textit{prior} predictive analysis, which is, we claim, a compulsory part of doing Bayesian data analysis.

\begin{figure}[t]
    \sidecaption[t]
    \includegraphics[scale=0.41]{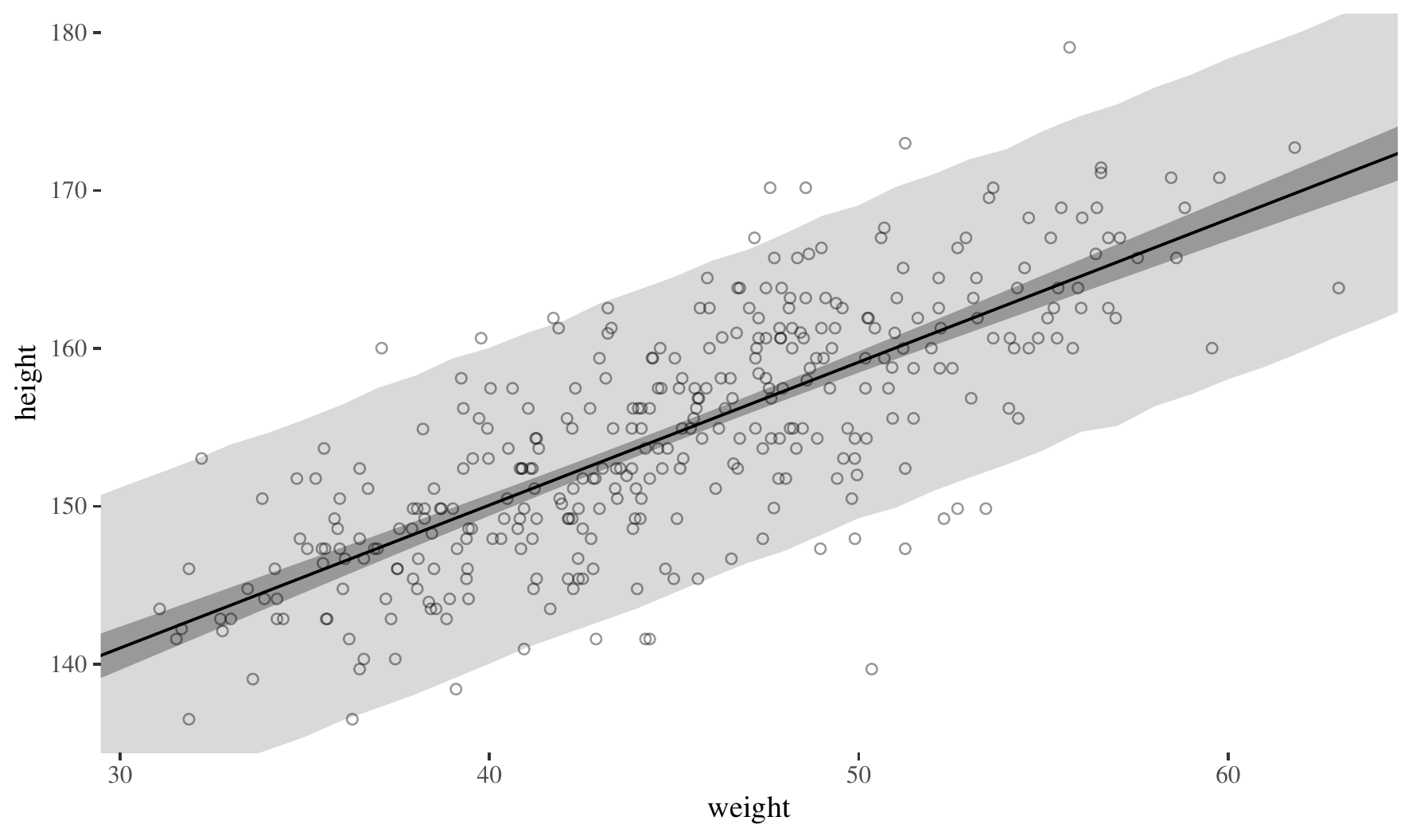}
    \caption{Height as a function of weight. The line represents the $\mu$, the dark shade the 95\% distribution of $\mu$, and the lighter shade the 95\% plausible region}
    \label{fig:plotted-est}
\end{figure}

For actually making inferences, i.e., determining the likely ranges of the parameters given our model, we will need data. We will make use of a data set found in the \texttt{rethinking} \textsf{R} package and a \textsf{R} Markdown script of our analysis can be downloaded.\footnote{\url{https://github.com/torkar/BDA_in_ESE}} After sampling, we will have a \textit{posterior} distribution, which is proportional to the likelihood and the prior distribution. In Fig.~\ref{fig:plotted-est}, we have plotted the empirical data set (circles) and the linear prediction (straight line). The narrow shaded interval is the 95\% distribution of $\mu$ (i.e., the exact values for many $\mu$ from the posterior), and the wider and lighter interval is the 95\% plausible region (i.e., 95\% of our $\mu$ should be found within that region).

Having a posterior predictive distribution we can now start to conduct various inferences, but we will stop here for now and instead, in Sect.~\ref{sec:case2}, present multiple ways we can make use of a posterior.

\begin{shaded}

To summarize, the three main steps of Bayesian model design and analysis are:

\begin{enumerate}
    \item Understand the data and the problem.
    \item Design a probability model (conduct model checking and iterate if the model needs to be revised) and sample from the posterior to conduct diagnosis.
    \item Conduct inference. That is, learn something about the population by using the posterior probability distribution, e.g., by conducting statistical tests or deriving estimates.
\end{enumerate}
\end{shaded}

The above is an iterative process, and in the last step, we also can change the parameters to see how they affect the outcome variable, i.e., to analyze the practical implications of different scenarios and thus assessing the practical significance of the results. The above steps will next be covered in a detailed case study within empirical software engineering.

\section{Case Study}\label{sec:case2}
Most would argue that to conduct estimations in software projects, one should not rely exclusively on expert opinion, but also on quantitative data collected in a more unbiased way. To this end, researchers have published studies making use of, among others, the International Software Benchmarking Standards Group's data repository (ISBSG).\footnote{\url{http://isbsg.org}} (For an overview and introduction to the data sets please see~\citep{hillSLW01isbsg}.) Their large collection of data sets include cost, size, and defect data from projects and sprints, which can be used for research, estimation, and prediction for and in future projects.

While the ISBSG data sets are typically cleaned and anonymized version of data sets collected from industrial projects they, like similar data sets in other collections, still exhibit many of the same characteristics that we can expect from actual projects in industry. For example, they have missing data, disparate quality in data collection procedures, and a large variety of data types. These are data quality issues we see also in \ESE{} research in general.

As we will later see in Sect.~\ref{sec:reading}, the dominant strategy to handle missing data in \ESE{} research is to merely remove cases that have missing data (listwise deletion). We believe that this strategy is sub-optimal and, generally speaking, not good for our research discipline. Even in cases when data can be classified according to the quality of the data collection procedure, as is the case with the ISBSG data sets, one sees that our community often chooses only to use a subset of data, classified to be of the highest quality (see, e.g.,~\citep{keung08dqr, liebchenS08dataquality} for recommendations, and~\citep{mittasPAA15lefteris} for an example where the authors use the recommendations). In short, we believe that data of low quality should be seen as better than no data at all, and the general rule of thumb should be never to throw away data. This is the context for our showcase and, in the following, we will both apply techniques for data imputation and conduct Bayesian data analysis on effort estimation data on the ISBSG data set collection. \footnote{A reproducibility package, making use of brms~\citep{brms} (with Stan~\citep{Stan}) written in \textsf{R}~\citep{R}, can be downloaded: \url{https://github.com/torkar/BDA_in_ESE}. The raw data can, however, not be downloaded due to copyright reasons. Please see \texttt{README.txt} in the repository for more information and what you need to do to access the raw ISBSG data.} As we will see, missing data can be naturally handled in Bayesian analysis and, thus, showcases one of the unique and pertinent strengths in an \ESE{} context.

\subsection{The Data and the Problem}\label{sec:case}
We will use the ISBSG Release 10 data set and set the dependent variable to \texttt{Effort}, i.e., the total number of person-hours to conduct a certain development task. According to, e.g.,~\citep{keung08dqr, liebchenS08dataquality} and the International Software Benchmarking Standards Group (ISBSG), the following pre-processing steps are appropriate:

\begin{enumerate}
    \item Only projects classified with data quality rating `A' are kept, and `B--D' are excluded.
    \item Only projects using IFPUG (unadjusted functional size measurement) should be kept. However, the data description clearly states that versions $\geq 4.0$ should not be compared to $<4.0$. Hence, we only use versions $\geq 4.0$.
    \item According to~\citet{keung08dqr}, some additional variables should be kept for compatibility with previous studies.
    \item Cases with missing values should be excluded. 
\end{enumerate}

The above leads to variables of interests as listed in Table~\ref{tbl:vars}, according to~\citet{mittasPAA15lefteris}. If we use the variables in Table~\ref{tbl:vars}, and follow the advice above, we will later see that we need to remove 3,895 projects out of the total 4,106 (close to 95\% of the projects). This seems wasteful but is the practice in our discipline, given the current standards and relevant recommendations.

\begin{table}
\centering
\caption{Variables of interest according to previous studies. The variable names that are underlined have been removed in this study as explained in Sects.~\ref{subsec:CA} and \ref{subsec:bias}. The variable name in bold was added as explained in Sect.~\ref{sec:case}}
\label{tbl:vars}
\begin{tabularx}{\columnwidth}{X X}
\hline
Name & Description \\
\hline
\underline{\texttt{AFP}}        & Adjusted function points\\
\texttt{Input}                  & Number of inputs\\
\texttt{Output}                 & Number of outputs\\
\texttt{Enquiry}                & Number of enquiries\\
\texttt{File}                   & Number of files\\
\texttt{Interface}              & Number of interfaces\\
\texttt{Added}                  & Number of added features\\
\texttt{Changed}                & Number of changed features\\
\underline{\texttt{Deleted}}    & Number of deleted features\\
\texttt{Effort}                 & Actual effort (person-hours)\\
\textbf{\texttt{DQR}}           & Data quality rating \\
\hline
\end{tabularx}
\end{table}

Imagine instead that we aim to keep as much data as possible, i.e., a data-greedy approach. Well, first of all, we should consider including all projects no matter the quality rating. After all, we can easily classify them differently in a statistical model and even investigate the difference between projects depending on the data quality rating. Hence, we decide to include all projects and mark them according to their data quality rating, i.e., \texttt{DQR}, no matter if they have missingness in them.

The next, crucial step before stating our model, is to analyze causality among our variables (in Table~\ref{tbl:vars}). While this was not needed in the simple, height-of-humans example above, it is essential in more complex situations when multiple variables measure entities that might be causally related. Since this is almost always the case in real-world \ESE{} research, it should be an essential step in building our statistical model. Otherwise, we risk that dependencies, e.g., correlations and collinearity, among variables might influence, and could potentially weaken our analysis. Below, we will then also directly analyze correlations between predictors, through an analysis of non-identifiability before we can then decide on our model we will also discuss how to handle the missing data and the sensitivity of our priors.

\subsubsection{Causal Analysis}\label{subsec:CA}
All predictors except \texttt{DQR}, seem to be raw and unadjusted measurements that are later used to calculate the adjusted function point (\texttt{AFP}). 

\begin{figure}
    \sidecaption
    \includegraphics[scale=.4]{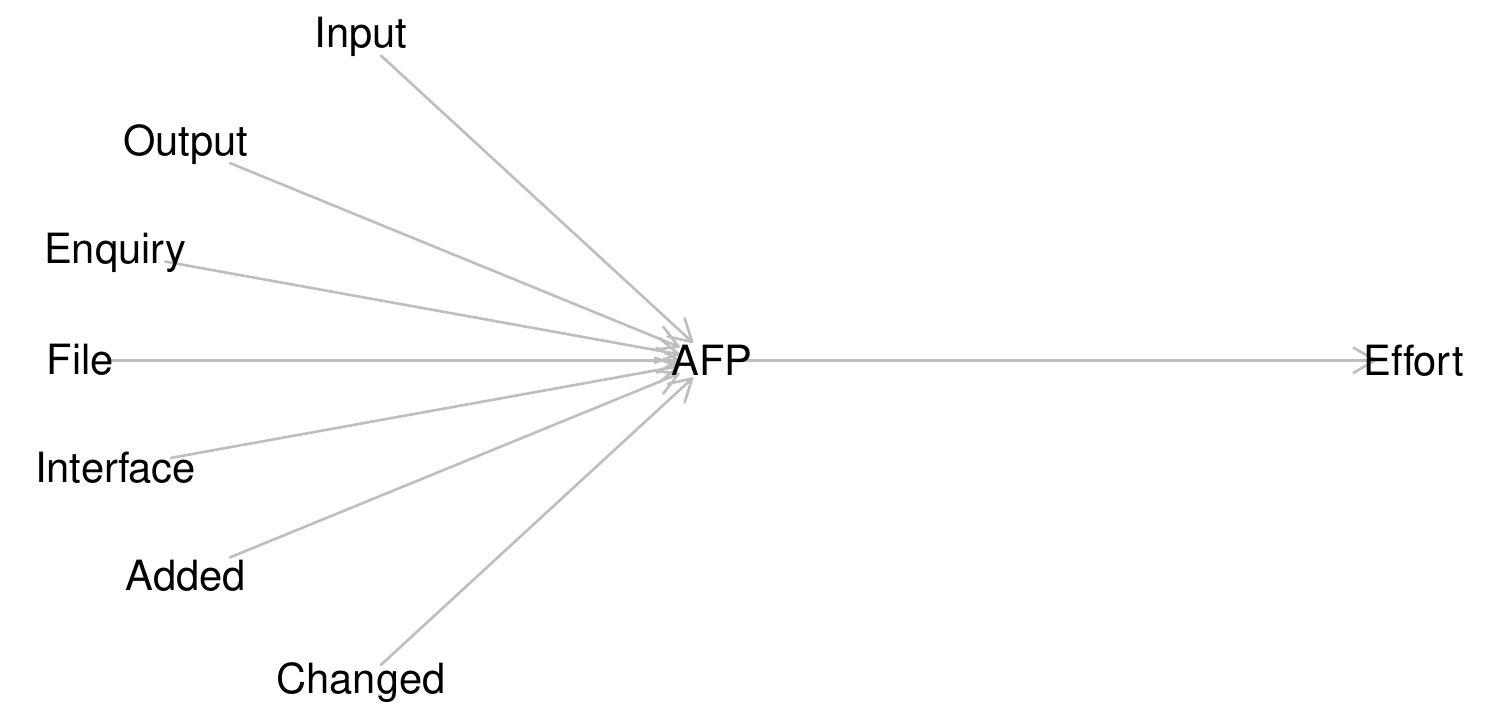}
    \caption{A directed acyclic graph of our scientific model}
    \label{fig:dag}
\end{figure}

Drawing our causal model (Fig.~\ref{fig:dag}) as a directed acyclic graph (DAG) shows something generally considered to be a pipe confounder (one of four types of relations in causal DAGs)~\citep{pearl09causality}.

In short, \texttt{AFP} mediates association between the other predictors and our outcome \texttt{Effort}, i.e., $\text{Effort} \bigCI \text{Input} | \text{AFP}$, or to put it differently, \texttt{Effort} is independent of \texttt{Input}, when conditioning on \texttt{AFP}.

We often worry about not having a predictor that we need for making good predictions (omitted variable bias). However, we do not often consider mistaken inferences because we rely on variables that are derived from other variables, i.e., post-treatment bias~\citep{mcelreath15statrethink}. (\citet{rosenbaum84post-treatment} calls this the concomitant variable bias.) In experimental studies, one would declare \texttt{AFP} to be a post-treatment variable, but the same nomenclature can be used in observational studies. To summarize, the above indicates that \texttt{AFP} should not be a predictor since it is derived from other, more basic, variables, so we will leave it out for now.

Note that causal analysis is not absolute in the sense that there is only one possible causal model that could be posited. Much scientific debate might be needed to argue for one or the other specific model and, thus, could lead to the exclusion of different sets of variables by different authors. As a consequence, this might lead different researchers to include or exclude different sets of variables and, thus, obtain different statistical results from separate analyses of the same data. However, in theory, this uncertainty and apparent subjectivity are still present with a traditional approach to statistical analysis, albeit hidden under the simplicity and familiarity of just applying a known statistical test.

\subsubsection{Identifying Non-Identifiability}\label{subsec:bias}
Strong correlations between variables are generally speaking a challenge when building statistical models. The model will make good predictions, but it will be harder to analyze and understand since the impression will be that variables that have a strong correlation, do not seem to have predictive power, while in fact, they have strong associations with the outcome~\cite[Sect.\ 5]{mcelreath15statrethink}. As a concrete example, if you want to build a model of a person's length, using the length of both her legs as separate predictors will not help the matter, i.e., adding another leg will not add predictive power to the model since it will correlate very strongly with the length of the first included leg. This type of multi-collinearity we generally want to avoid in statistical models.

Traditionally, there are two ways one can investigate this: Examining a pairs plot where all combinations of parameters and their correlations are visualized, or check if the matrix of predictor values is a full rank matrix, and thus identify non-identifiability that way.\footnote{In short, different values of the parameters must generate different probability distributions of the observable variables. Otherwise we face various degrees of non-identifiability, i.e., essentially that (too) many parameter combinations could lead to the same observations.} 

The latter, matrix-based approach consists of declaring a model $y = \beta_1 x_1 + \ldots + \beta_n x_n$, using the data, i.e., the values of the predictor variables, $x_i, \ldots, x_n$ as a matrix $\mathbb{A}$, and decompose it into a product $\mathbb{A} = \mathbb{QR}$ of an orthogonal matrix $\mathbb{Q}$, and an upper triangular matrix $\mathbb{R}$. By analyzing the diagonal of the matrix $\mathbb{R}$ a threshold value of $\lvert d_{ij} \rvert< 0.1$ along the diagonal is then often used to declare a variable as unsuited for inclusion in a model. Often something like $1e^{-12}$---quasi-zero for a computer---could be used, but generally speaking anything below 1.0 has traditionally been excluded. The argument is that if it is below 1.0, then the variable would provide very little additional value to the model and should thus not be included.

Identifying non-identifiability for our data set from Table~\ref{tbl:vars} clearly indicates that \texttt{Deleted} should be a candidate for removal ($\lvert d_{\text{Deleted}} \rvert = 9e^{-12}$). It might feel strange to remove predictors when our ultimate goal is to use as much data as possible. However, to build a statistical model that is sane, has good out of sample prediction, and is understandable, a trade-off is needed. Here we argue that the bulk of work should be done before we design our statistical model, to make use of the missing data techniques available to us.

\subsubsection{Missing Data Analysis}\label{subsec:MDA}
When data is missing from cases of our data sets the most common solutions is to either delete such cases or impute, i.e., `guess', based on the values we do have. \citet{rubin87imp} has shown that very often 3--5 imputations are enough (the complete dataset is imputed fully 3--5 times) and that the relative efficiency of an estimate based on $m$ imputations is approximately:

\begin{equation*}
    \text{Relative Efficiency} \approx (1 + \frac{\gamma}{m})^{-1}
\end{equation*}
\noindent
where $\gamma$ is the fraction of missing information. The relative efficiency in this case refers to using the finite $m$ imputation estimator instead of the infinite number for the fully efficient imputation.

As an example, consider that we have 20\% missing information in a variable ($\gamma=0.2$), given $m=5$, we have reached a relative efficiency of approximately 96\%. Setting $m=10$ we reach 98\%. By doubling the computational effort, we have only a slight gain in relative efficiency. On the other hand, we have lots of computing power at our hands nowadays. However, more recently, we have seen that other recommendations for handling missing data have been presented.

\citet{bodner08mi} and \citet{whiteRW11im} showed through simulations and by analytically deriving Monte Carlo errors, respectively, that the general rule of thumb should be $m = \gamma \times 100$, i.e., if a variable has 40\% missing data ($\gamma = 0.4$) we should set $m=40$ (using Rubin's efficiency estimate this would mean 92.6\%$\to$99.0\%).

As will be evident, we will take the more conservative approach (i.e., $m = \gamma \times 100$), when we present the model implementation in Sect.~\ref{subsec:BDA}.

Before we continue with the next section, it might be worthwhile to note that missingness in a Bayesian framework can be done in different ways. Either we conduct multiple imputations to derive uncertainty for all parameters, including our missing data. This is the path we have chosen here. The other approach would be to design a model of all data, including the missingness mechanism. 

\begin{shaded}
To model the missingness mechanism can be more involved and requires us to be very explicit about how our missingness occurred. At the best of times, this is a challenging task. We could instead argue that using the first approach, as we do here, shows the strengths of the Bayesian approach, since it easily can make use of various techniques, to handle missing data, in a coherent and principled way.
\end{shaded}

\subsubsection{Sensitivity Analysis of Priors}\label{subsec:sensitivity}
Our analysis so far, to summarize, indicates that we should use seven predictors and one group variable (quality level of the project, in ISBSG terms), to predict one outcome variable (Effort). We can already now assume that we will most likely use a likelihood (i.e., our assumptions regarding the data generative process) that is based on counts (\texttt{Effort} is after all a count, the number of hours, going from zero to infinity).

Thus, we plan to use a generalized linear model, with a link function that translates between the linear predictor value and the mean of the distribution function. In the case of count distributions, such as Poisson, it is customary to use a $\log$ link function, i.e., a parameter's value is the exponentiation of the linear model. However, when setting priors for parameters and using a link function, unexpected things can happen, and the priors might not have the effect one would expect. To this end, we should always do prior predictive simulations, i.e., a sensitivity analysis of how different settings of the priors affect the predicted variable.

The description of the data set indicates that approximately 20\% of the projects have more than 20 people in the team. If we assume, roughly 1,700 h\slash year for an individual, having 60 people in a team sums up to approximately 100,000 person-hours per year. Let us now assume that this is the maximum value for our outcome variable \texttt{Effort}. Random sampling from $e^{\mathcal{N}(5,4)}$ provides us with $\bar{x}\approx 208,000$ (we use the exponential since we assume a $\log$ link function) indicating that this could be an acceptable prior for the intercept $\alpha$.

We arrived at the values 5 and 4 above by starting from typical default values such as assuming $\alpha$ to be $\mathcal{N}(0,10)$ (not uncommon as a default choice in, e.g., Poisson regressions), this would lead to $\bar{x} = 8 \times 10^{11}$ hours of work effort for an average project. This would correspond to close to half a billion people working on the project for one year. Thus we should use non-default priors to adapt the priors better so that they do not (often) give absurd values.

\begin{figure}
\centering
\subfloat[$\beta$ priors $\mathcal{N}(0,10)$]{\includegraphics[scale=0.15]{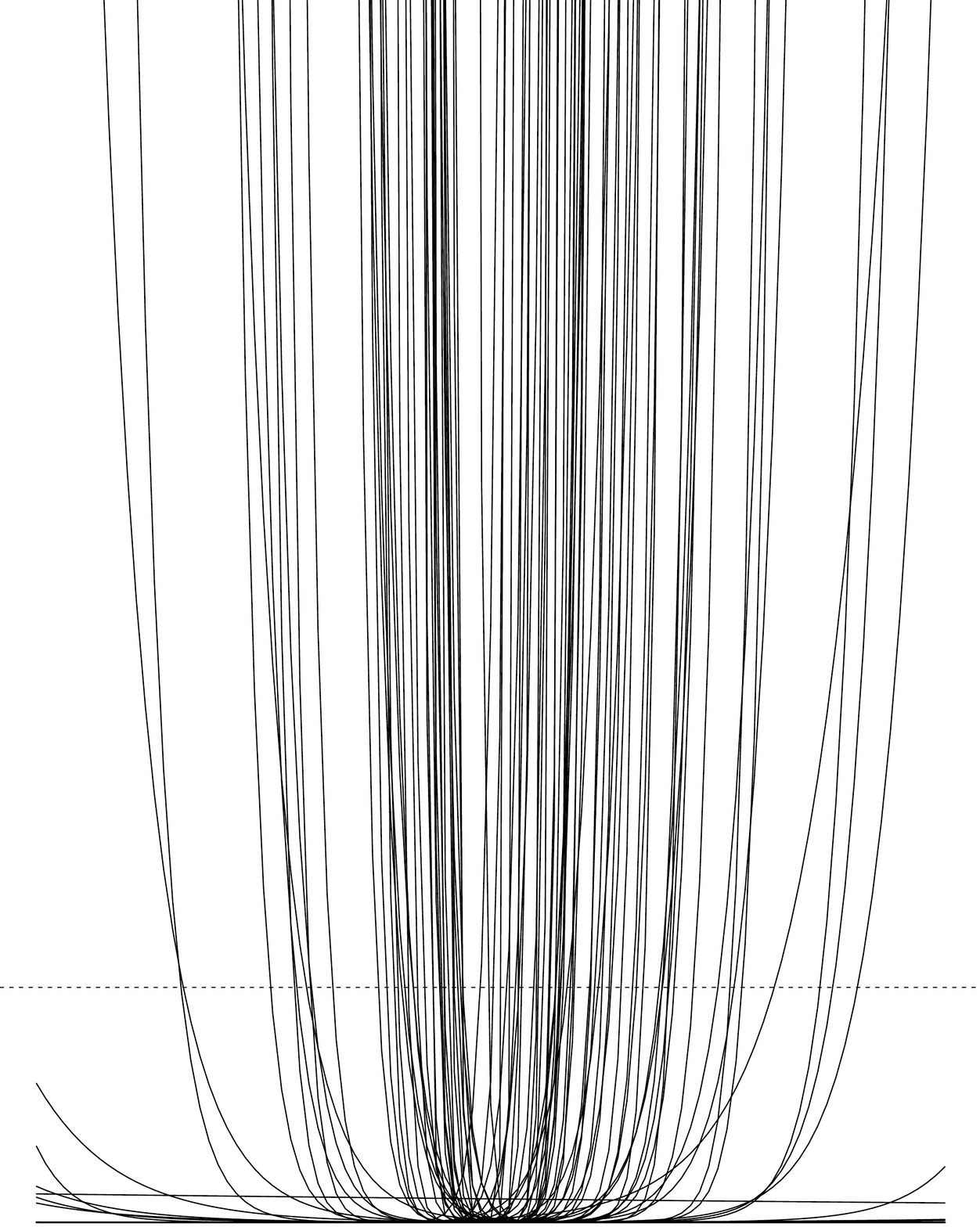}}%
\label{fig:broad}
\hfil
\subfloat[$\beta$ priors $\mathcal{N}(0,0.25)$]{\includegraphics[scale=0.15]{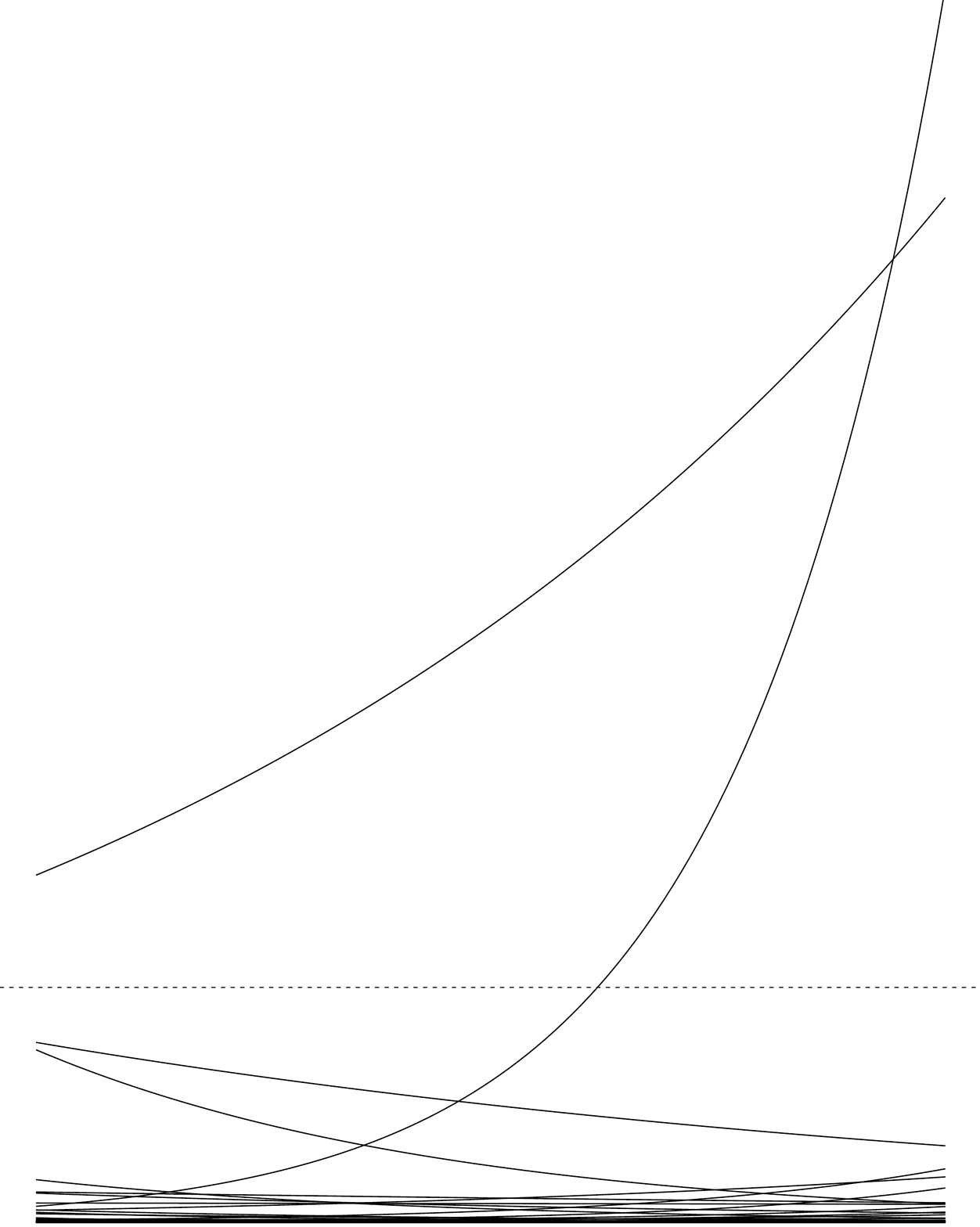}}%
\label{fig:tight}

\caption{Prior predictive simulation of broad and informative priors, respectively. The $x$-axes are $z$-scores, while the $y$-axes represents our outcome variable \texttt{Effort}. The dashed horizontal line corresponds to our assumed maximum value for our outcome variable, and is the only value of interest in this case. We have plotted 100 simulations with the intercept $\mathcal{N}(5,4)$ and our seven priors for the respective $\beta$ parameters}
\label{fiG:prior_pred}
\end{figure}

To assess the impact of very broad priors like $\mathcal{N}(0, 10)$ for our seven parameters, we thus, iteratively, compared their usage to that of other, narrower priors like $\mathcal{N}(0, 0.25)$. Furthermore, assuming a $\log$ link function, the additive effects of the seven priors for our $\beta$ parameters would, on a normalized scale, correspond to $\mathcal{N}(0, (10 \times 7)^2)$ and $\mathcal{N}(0, (0.25 \times 7)^2)$, respectively. As is evident from Fig.~\ref{fiG:prior_pred} (a), we have a massive emphasis on extremely high $y$-values (we would require the world's total population to work in a project for this to happen). Now compare (a) with (b). We still allow extremely large $y$-values (up to $640\times10^6$!), but the emphasis is now a bit more realistic.

To summarize, prior predictive simulations indicate that setting $\mathcal{N}(5, 4)$ and $\mathcal{N}(0, 0.25)$ on $\alpha$ and $\beta$, respectively, allows us to delimit the multidimensional Gaussian space of possible parameter values, while still not remove the probability of extreme values altogether. If we would still be uncertain, one could have conducted even more prior predictive simulations~\citep{SimpsonRMRS14}. The prior knowledge we took with us when doing this analysis, was that it was not likely that very many projects had billions of people involved. One could have taken a more conservative approach and claim that it is not likely that we have millions of people in our projects; however, as we will see, Hamiltonian Monte Carlo will handle these priors well given the available data.

\begin{shaded}
One should always conduct prior sensitivity analysis (prior predictive simulation) before making use of the available data. There is always \textit{some} prior knowledge one can use!
\end{shaded}

\subsection{Design of Model and Diagnosis}\label{sss:design}
Based on our initial analysis, we have a much clearer picture of which variables to include and the overall sensitivity of our priors.
For our model one could imagine using a Poisson likelihood, that is, we have a count ($\text{\texttt{Effort}}_{0 \to \infty}$), which we can model binomial events for when the trials $N$ are very large and the probability $p$ small. However, that would be a mistake. %rf: Reader's are not likely to get this level of statistical detail without it being connected to their Se knowledge of the situation. What would N and p refer to in Se terms? Why would it be a mistake. Do we even need to mention this alternative given we assumed a GLM with poisson distribution above?
%rto this was more as an intro to the section to remind the reader that we assumed a Poisson at first. But as we will see this does not hold. It's natural that we do this step by step and iterate.

In any analysis, it is important first to get to know the actual data. Thus, let us look at some descriptive statistics of the data (taking into account the pre-processing steps previously introduced).\footnote{For all data sets we use single quotes to emphasize the names, e.g., `A\_clean', while we print out variable names in \texttt{verbatim}.} Some issues catch the eye in Table~\ref{tbl:descr_data}. First, \texttt{Effort} has a max value of 645,694 (three times larger than the mean for our priors). Second, the medians are consistently lower than the means (in one case the median is zero) indicating positive skewness. Third, not visible in the table, \texttt{Effort}, compared to the predictors, contain no zeros (indicating that we do not need to consider zero-inflated or hurdle models~\citep{huPNzip}). Finally, the mean and the variance for our outcome variable are very different (the variance is approximately 70,000 times larger than the mean).

\begin{table*}
\centering
\caption{Descriptive statistics of our predictors and the outcome variable \texttt{Effort} using all data available to us (i.e., 4,106 projects).  after conducting the pre-processing steps in Sect.~\ref{sec:case}. From left to right: Name, mean, median, max, min, and sample variance (with removed NAs). All numbers rounded to the nearest integer}
\label{tbl:descr_data}
\begin{tabularx}{\textwidth}{X S S S S S S}
\hline
\multicolumn{1}{l}{Variable} & \multicolumn{1}{c}{$\bar{x}$} & \multicolumn{1}{c}{$\tilde{x}$} & \multicolumn{1}{c}{$\text{max}(x)$} & \multicolumn{1}{c}{$\text{min}(x)$} & \multicolumn{1}{c}{$s^2$}\\
\hline
\texttt{Input}       & 143 & 56 & 9404 & 0 & 144167\\
\texttt{Output}      & 125 & 47 & 3653 & 0 & 70522\\
\texttt{Enquiry}     & 74 & 27 & 2886 & 0 & 22318\\
\texttt{File}        & 118 & 43 & 10821 & 0 & 137172\\
\texttt{Interface}   & 39 & 10 & 1572 & 0 & 11082\\
\texttt{Added}       & 357 & 142 & 15121 & 0 & 591300\\
\texttt{Changed}     & 128 & 0 & 18357 & 0 & 344677\\
\texttt{Effort}      & 5384 & 1828 & 645694 & 4 & 391631309\\
\hline
\end{tabularx}
\end{table*}

Concerning the latter issue, a Poisson likelihood assumes the mean and the variance be approximately equal. This allows us to use the negative binomial, known as the Gamma–Poisson (mixture) distribution, as our likelihood, i.e., a continuous mixture model where we assume each Poisson count observation has its own rate. However, since we are still using a Poisson model, in essence, one could claim that we do not redo the sensitivity analysis.

To summarize our findings so far we can now formulate our model:

{\footnotesize
\begin{IEEEeqnarray}{rCl}
\text{Effort}_i & \sim & \text{Gamma-Poisson}(\lambda_i, \phi_i)\nonumber\\
\log(\lambda_i) & \sim & \alpha + \beta_{\text{Input}} \times \text{Input} + \beta_{\text{Output}} \times \text{Output} + \beta_{\text{Enquiry}} \times \text{Enquiry}\nonumber\\ 
&& +\: \beta_{\text{File}} \times \text{File} + \beta_{\text{Interface}} \times \text{Interface} + \beta_{\text{Added}} \times \text{Added}\nonumber\\ && +\: \beta_{\text{Changed}} \times \text{Changed} + \alpha_{\text{DQR}[i]}\nonumber\\
\alpha & \sim & \mathcal{N}(5,4)\nonumber\\
\beta_{1, \ldots, 7} & \sim & \mathcal{N}(0,0.25)\nonumber\\
\alpha_{\text{DQR}} & \sim & \mathcal{N}(0, \sigma)\nonumber\\
\sigma & \sim & \text{HalfCauchy}(0,1)\nonumber\\
\log(\phi_i) & \sim & \text{Gamma}(0.5, 0.5)\nonumber
\end{IEEEeqnarray}
}

We model each observation from a negative-binomial (Gamma-Poisson) distribution, with a failure rate $\lambda$ and shape $\phi$. We then use a $\log$ link for our linear model $\lambda$ where we include an intercept $\alpha$ and parameters ($\beta$) for all predictors. 

We also add varying intercepts in the form of our \texttt{DQR} variable. The idea is that each data quality rating should be treated uniquely by allowing us to estimate $\alpha$ for each rating, i.e., each \texttt{DQR} will have its own intercept. This will enable us to see if there is an overall difference between projects judged to have different quality ratings.

Finally, we set the aforementioned priors on our $\beta$ parameters (but we use $\mathcal{N}(0, \sigma)$ for our unique intercepts, to separately estimate $\sigma$ for each level of \texttt{DQR}). We also set $\text{HalfCauchy}(0, 1)$ and $\text{Gamma}(0.5, 0.5)$ for $\sigma$ and $\phi$, respectively. Both of these priors are regularizing priors common for these types of parameters.\footnote{Please see here for prior choice recommendations: \url{https://goo.gl/fx2F7V}.}

\subsubsection{Using the Model}\label{subsec:BDA}
In the previous sections, we presented our statistical model with assumptions. In this section, we will make use of it in two ways: sampling with complete data and imputed data. However, before we begin, Table~\ref{tab:descr_samples} describes the data sets we will use.

\begin{table}
    \centering
    \caption{Data sets used. The `A*' and `AD*' categories are of different dimensions due to our index variable, \texttt{DQR}, added to the `AD*' sets. From left to right. Name of data set, number of projects (rows), number of NAs, percentage of NAs, and number of zeros}
    \label{tab:descr_samples}
    \begin{tabularx}{\columnwidth}{X S S S S}
    \hline
 \multicolumn{1}{l}{Name} & \multicolumn{1}{r}{\# projects} & \multicolumn{1}{r}{\# NAs} & \multicolumn{1}{r}{\% NAs} & \multicolumn{1}{r}{\# zeros}\\
    \hline
    `A'                   & 501   & 2109  & 23.8     & 316\\ % d_ifpug_A
    `A\_clean'            & 214   & 0     & 0        & 316\\ % d_ifpug_A_clean
    `AD'                  & 1689  & 8507  & 19.8     & 736\\ % d_ifpug_AD
    `AD\_clean'           & 494   & 0     & 0        & 727\\ % d_ifpug_AD_clean
    \hline
    \end{tabularx}
\end{table}

The data sets are divided into two categories. First, we have data sets that only take into account projects classified as having the highest quality rating (`A') and data sets where we use all four quality ratings (`AD'); taking into account the pre-processing steps in Sect.~\ref{sec:case}.

First, we have subsets with NAs (`A' and `AD') and, second, subsets where all original NAs are removed (`*\_clean'). The logic to use these data sets is that we want to use as much data as possible (but we pay the price of missing data), and removing all NAs is, as we have discussed, not uncommon.

One could also imagine having subsets where all zeros are assumed to be NAs, but that would be a bit too conservative assumption in our opinion, and we leave it to the reader to try out such a scenario.

If one would like to compare our data sets with what is commonly seen in literature then, taking into account that we have a more restrictive view on which IFPUG versions are included, `A\_clean' would be the most similar data set (e.g., \citet{mittasPAA15lefteris} report using 501 projects, while we end up with 214, using our more restrictive subset). However, we are more interested in the cases where we have larger data sets, together with missing data, and comparing these with, e.g., `A\_clean'.

\paragraph{Missing Data Imputation}
Summarizing missing data (Table~\ref{tab:md}) shows that the missingness is multivariate (there is missing data in more than one variable), connected (the second row with data indicates that we have 214 rows that are complete, i.e., no data is missing for these rows), and non-monotone (we have zeros spread out within all the ones, i.e., there is no monotonicity). Generally speaking, this indicates that data imputation is possible (in particular, connectivity is an essential part of missing data imputation).

\begin{table}
    \centering
    \caption{Missingness of missing data. Top row lists each variable. Bottom row the number of missing entries per variable. First column, the frequency of each pattern. Last column, number of missing entries per pattern}
    \label{tab:md}
    \begin{tabularx}{\textwidth}{p{2cm} rrrrrrrrr}
    
&    Effort &Added &Input &Output &Enquiry &File &Interface &Changed & \\    
\hline
\textbf{Freq.} &&&&&&&&& \# \textbf{missing entries}\\
214 &     1   &  1  &   1  &    1   &    1  &  1    &     1 &   1  &  0 \\
14  &     1  &   1   &  1  &    1   &    1   & 1    &     1 &    0  &  1 \\
8   &     1  &   1  &   1  &    1   &    1  &  1    &     0 &      1 &   1 \\
2   &     1  &   1  &   1  &    1   &    1  &  1    &     0 &      0 &   2 \\
1   &     1  &   1  &   1  &    1   &    1  &  0    &     0 &      1 &   2 \\
2   &     1  &   1  &   1  &    1   &    0  &  1    &     1 &      0 &   2 \\
1   &     1  &   1  &   1  &    1   &    0  &  0    &     1 &      0 &   3 \\
3   &     1  &   1  &   1  &    0   &    1  &  1    &     1 &      0 &   2 \\
1   &     1  &   1  &   1  &    0   &    1  &  1    &     0 &      0 &   3 \\
1   &     1  &   1  &   1  &    0   &    1  &  0    &     0 &      1 &   3 \\
2   &     1  &   1  &   0  &    1   &    0  &  0    &     1 &      0 &   4 \\
5   &     1  &   1  &   0  &    0   &    0  &  0    &     0 &      1 &   5 \\
4   &     1  &   1  &   0  &    0   &    0  &  0    &     0 &      0 &   6 \\
1   &     1  &   0  &   1  &    1   &    1  &  1    &     1 &      1 &   1 \\
1   &     1  &   0  &   1  &    1   &    1  &  0    &     0 &      1 &   3 \\
84  &     1  &   0  &   0  &    0   &    0  &  0    &     0 &      0 &   7 \\
157  &    0  &   0  &   0  &    0   &    0  &  0    &     0 &      0 &   8 \\
\hline
\textbf{Missingness per variable} &   157  & 243  & 252  &  255   &  255 & 256     &  264  &   270 & $\mathbf{\sum}$ \textbf{1952} 
    \end{tabularx}
\end{table}

\begin{figure}[t]
    \sidecaption[t]
    \includegraphics[scale=0.45]{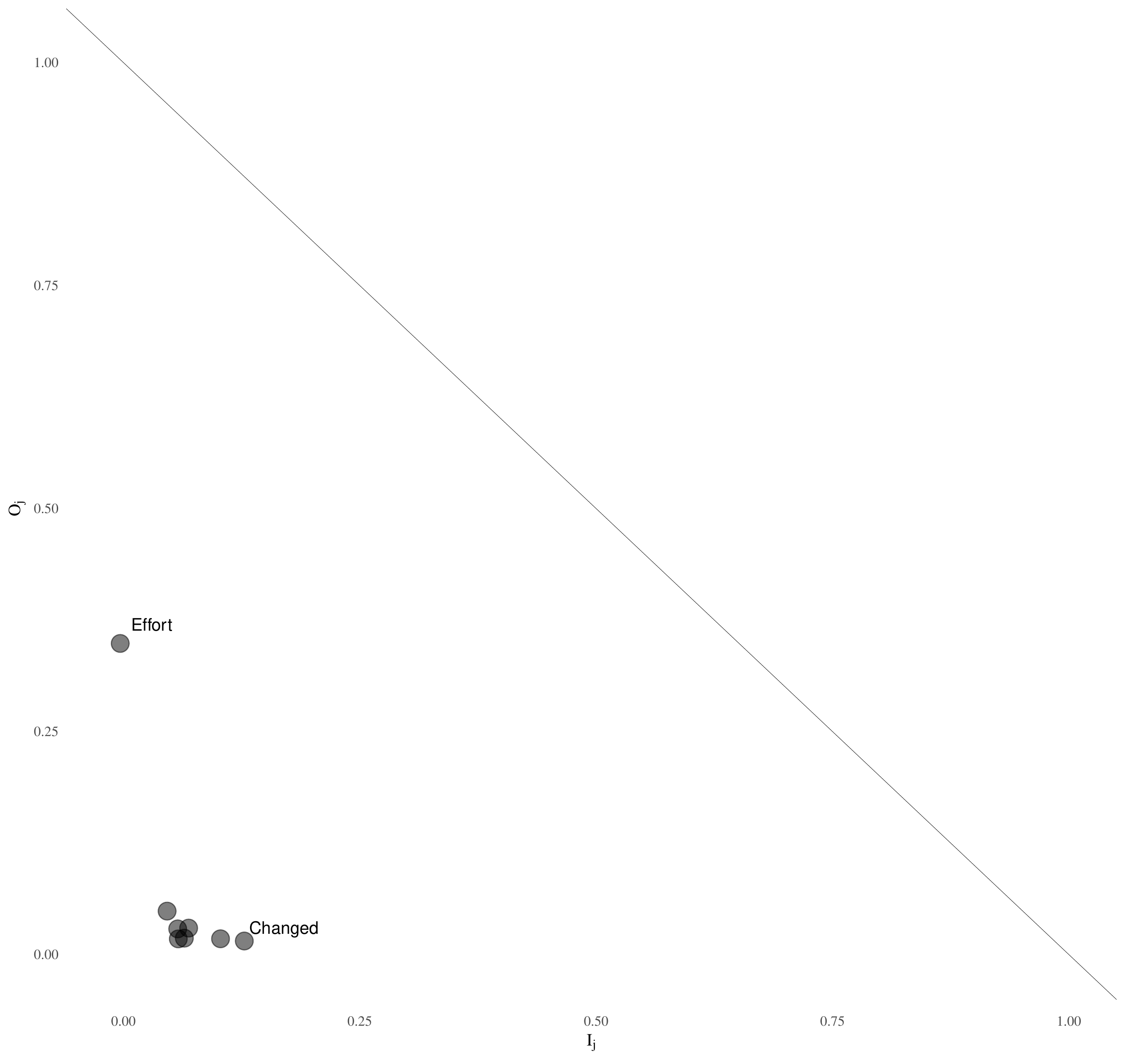}
    \caption{Outflux versus influx of data set `A' as described in Sect.~\ref{tab:descr_samples}. (A small degree of random noise, `jitter', was added to the plot to make it more readable.)}
    \label{fig:flux}
\end{figure}

\citet{vanBuuren07mice} recommends that one calculates each variable's influx and outflux and plots them. Influx ($I$) is defined as the number of variable pairs with $Y_j$ missing and $Y_k$ observed, divided by the total number of \textit{observed} data cells while, in the case of outflux ($O$), we instead divide by the total number of \textit{incomplete} data cells. In short, a completely observed variable gives $I_j = 0$, while the opposite holds for $O_j$. If one has two variables with the same degree of missing data, then the variable with the highest $O_j$ is potentially more useful for imputation purposes. Examining Fig.~\ref{fig:flux}, \texttt{Effort} and \texttt{Change} have the highest $O_j$ and $I_j$. To summarize, \texttt{Effort} will be the most influential variable for imputation, while \texttt{Change} will be the easiest variable to impute. This is worth keeping in mind later when we analyze the results.

To conclude, we will create multiple imputations (replacement values) for multivariate missing data, based on Fully Conditional Specification~\citep{vanBuuren07mice}. Each incomplete variable is imputed by a separate model using predictive mean matching (numeric data), or proportional odds model\slash ordered $\logit$ model (factor data with \textgreater 2 ordered levels)~\citep{rubin86mm, vanBuuren07mice}. Concerning predictive mean matching, the assumption is that the missingness follows approximately the same distribution as the data, but the variability between the imputations over repeated draws reflects the uncertainty of the actual value.

We will approach this conservatively and follow the latest guidelines, as already discussed in Sect.~\ref{subsec:MDA}, and hence set the number of imputations $m = 25$ (see Table~\ref{tab:descr_samples}) since we have approximately 25\% missingness in certain variables.

\subsubsection{Diagnostics}\label{subsub:diag}
In this section, we will first present some diagnostics from the Hamiltonian Monte Carlo sampling we conducted.

\begin{figure*}
\centering
\subfloat[]{\includegraphics[scale=0.35]{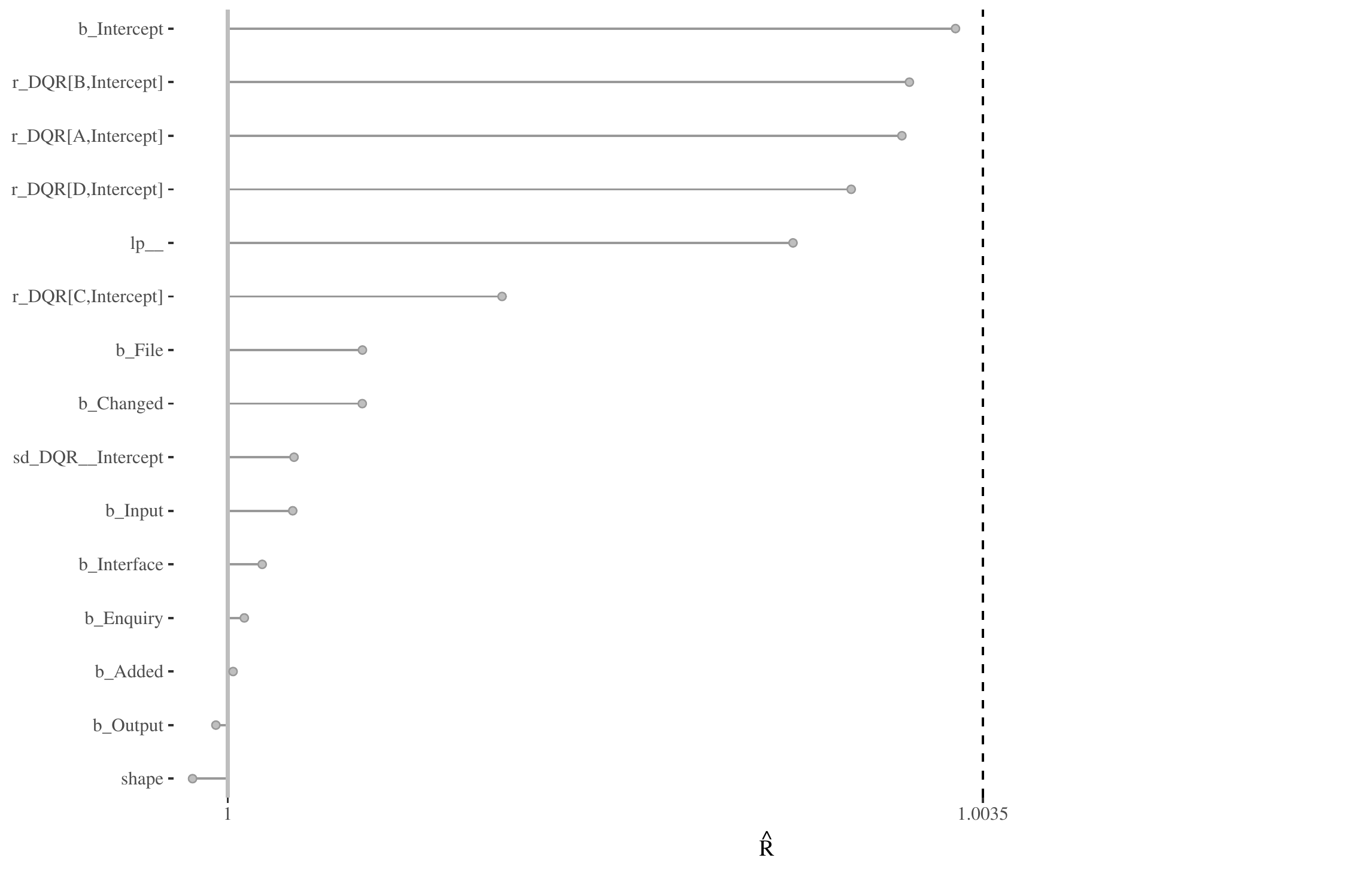}%
\label{fig:rhat-case2}}
\vfil
\subfloat[]{\includegraphics[scale=0.35]{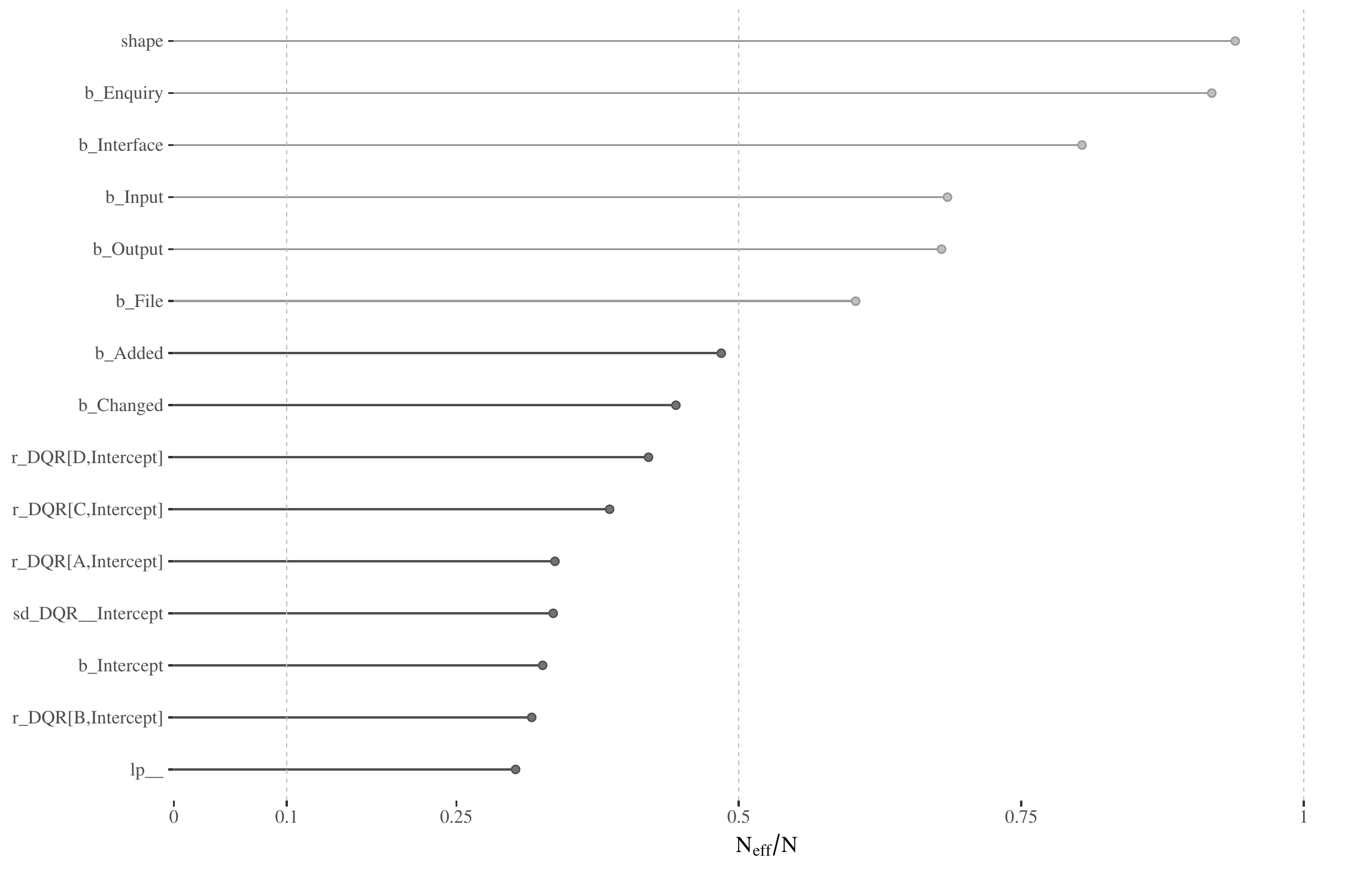}%
\label{fig:neff_case2}}
\caption{$\widehat{R}$ values for each parameter (a) and effective sample sizes (b). For $\widehat{R}$ (a), generally speaking, any values above 1.01 are not appropriate and indicates that one or more chains are biased. Concerning effective sample sizes (b), anything below 0.2 is generally speaking a warning sign of a misspecified model. Here we used the `AD\_clean' data set}
\label{fig:diag}
\end{figure*}

First, the ratio of the average variance of draws within each chain to the variance of the pooled draws across chains is an estimate we can use to see how well our chains have diverged towards a common posterior. This is measured by $\widehat{R}$ and generally speaking $\widehat{R}$ should go towards 1.00, and anything above 1.01 should be a clear warning sign of bias. In our case, for all sampling conducted, $\widehat{R}$ was consistently low (see Fig.~\ref{fig:rhat-case2} for one example where we used the `AD\_clean' data set).

Second, the effective sampling was consistently high, i.e., $\text{n}_{\text{eff}} \gg 0.2$. As an example, if we have 2,000 samples from each chain, the default is then to throw away the first 1,000 as warmup samples. But then we use four chains, ending up with $1,000 \times 4$ samples. This means that we should not, in our example, have less than 400 samples for a parameter (see Fig.~\ref{fig:neff_case2}).

\begin{figure}[t]
    \sidecaption[t]
    \includegraphics[scale=0.32]{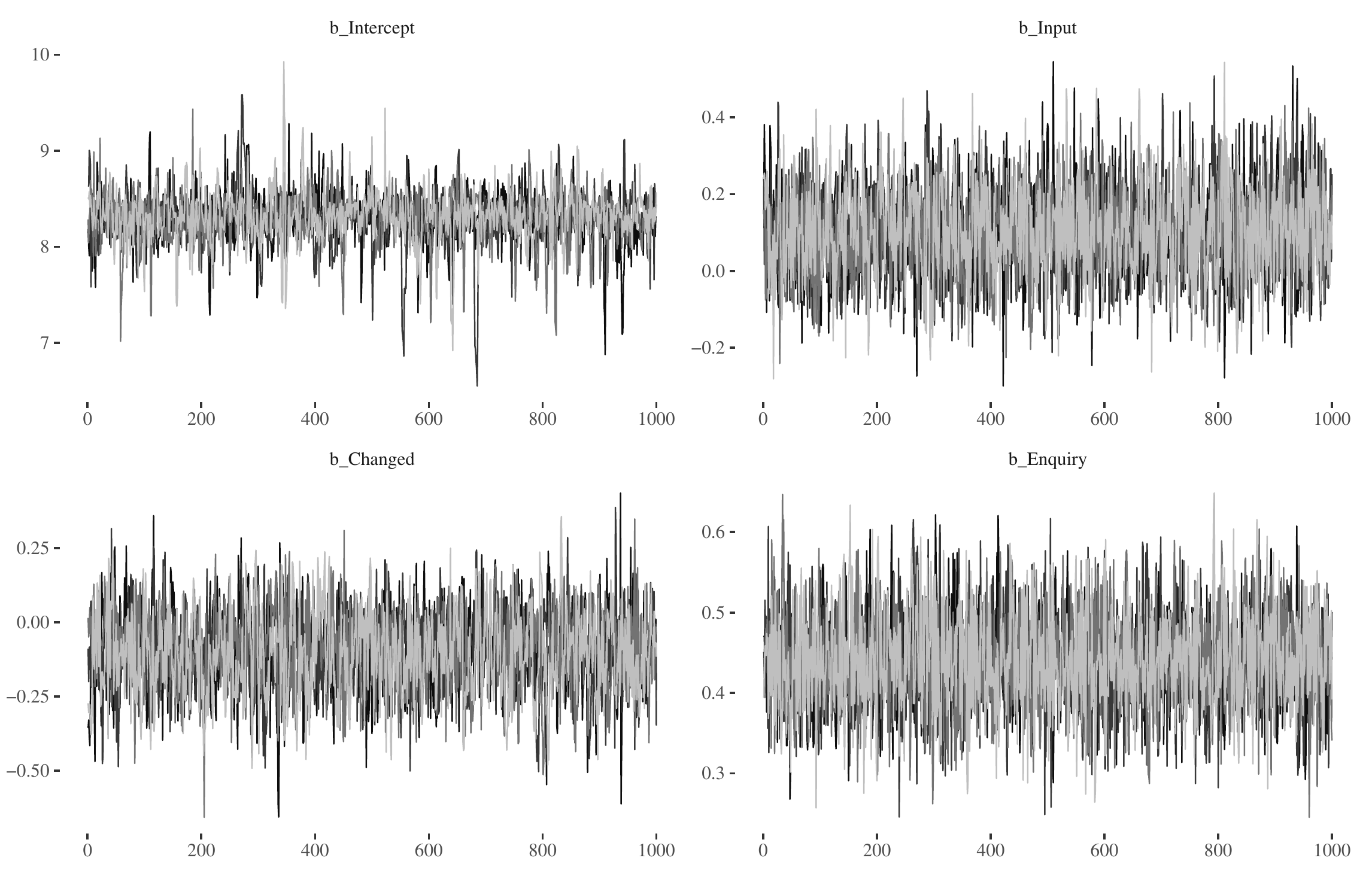}
    \caption{Trace plots of four parameters using the `AD\_clean' data set}
    \label{fig:trace-case2}
\end{figure}

\begin{figure}[t]
    \sidecaption[t]
    \includegraphics[scale=.32]{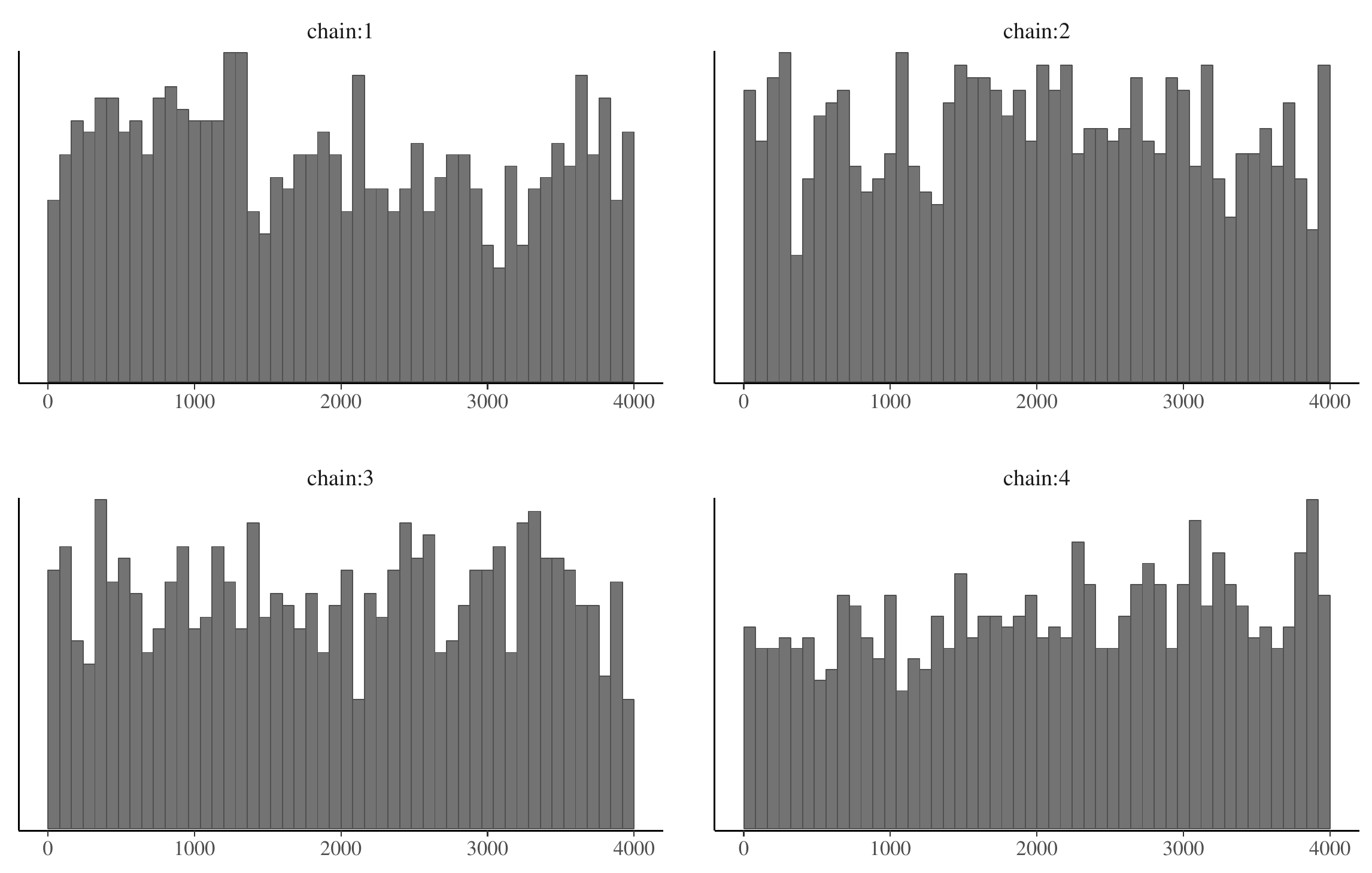}
    \caption{Rank plots of the four chains from the $\beta_{\text{Intercept}}$ sampling. The chains should be close to uniform. Here we see a slight dip at the start of Chain 1 and the end of Chain 3. These dips can be much more exaggerated (e.g., no samples collected at all) and then there would be reasons to worry}
    \label{fig:rank_case2}
\end{figure}

Third, the visual inspection revealed that the chains seemed to mix well (hairy caterpillar ocular test). It should look messy, tight, and mixed (Fig.~\ref{fig:trace-case2}). If we are a bit hesitant concerning the mixture, and in particular the sampling conducted at the tails, once could also use rank plots~\citep{vehtariGSCB19diag}. If we investigate Fig.~\ref{fig:trace-case2} we see a clear difference between $\beta_{\text{Intercept}}$ and the other parameters. Using rank plots for the chains for $\beta_{\text{Intercept}}$, Fig.~\ref{fig:rank_case2}, provides a better view. We can see that there is a dip at the start of Chain 1 and the end of Chain 3, but it still indicates that the chain managed to sample quite well at the tails.

\begin{figure}[t]
    \sidecaption[t]
    \includegraphics[scale=0.32]{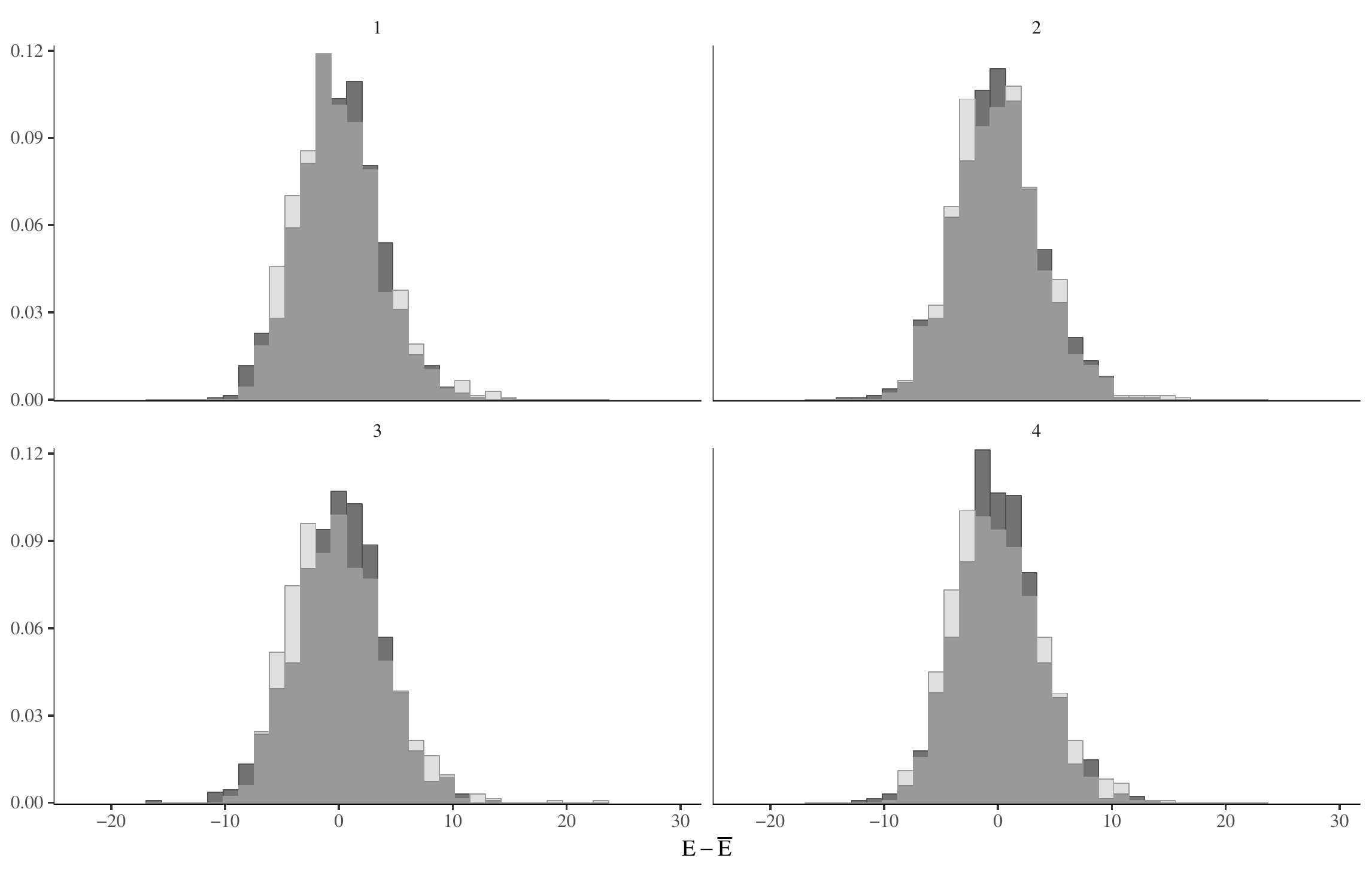}
    \caption{Comparisons of the energy transition density, $\pi_E$, and the marginal energy distribution, $\pi_{\Delta E}$ (light and dark gray,  respectively). A significant overlap is visible}
    \label{fig:bfmi}
\end{figure}

Finally, the Bayesian fraction of missing information, another diagnostic, shows a significant overlap between the energy transition density, $\pi_E$, and the marginal energy distribution, $\pi_{\Delta E}$ (Fig.~\ref{fig:bfmi}). When the two distributions are well-matched, the random walk will explore the marginal energy distribution efficiently~\citep{betancourt17hmc}.

\begin{figure}[t]
    \sidecaption[t]
    \includegraphics[scale=0.4]{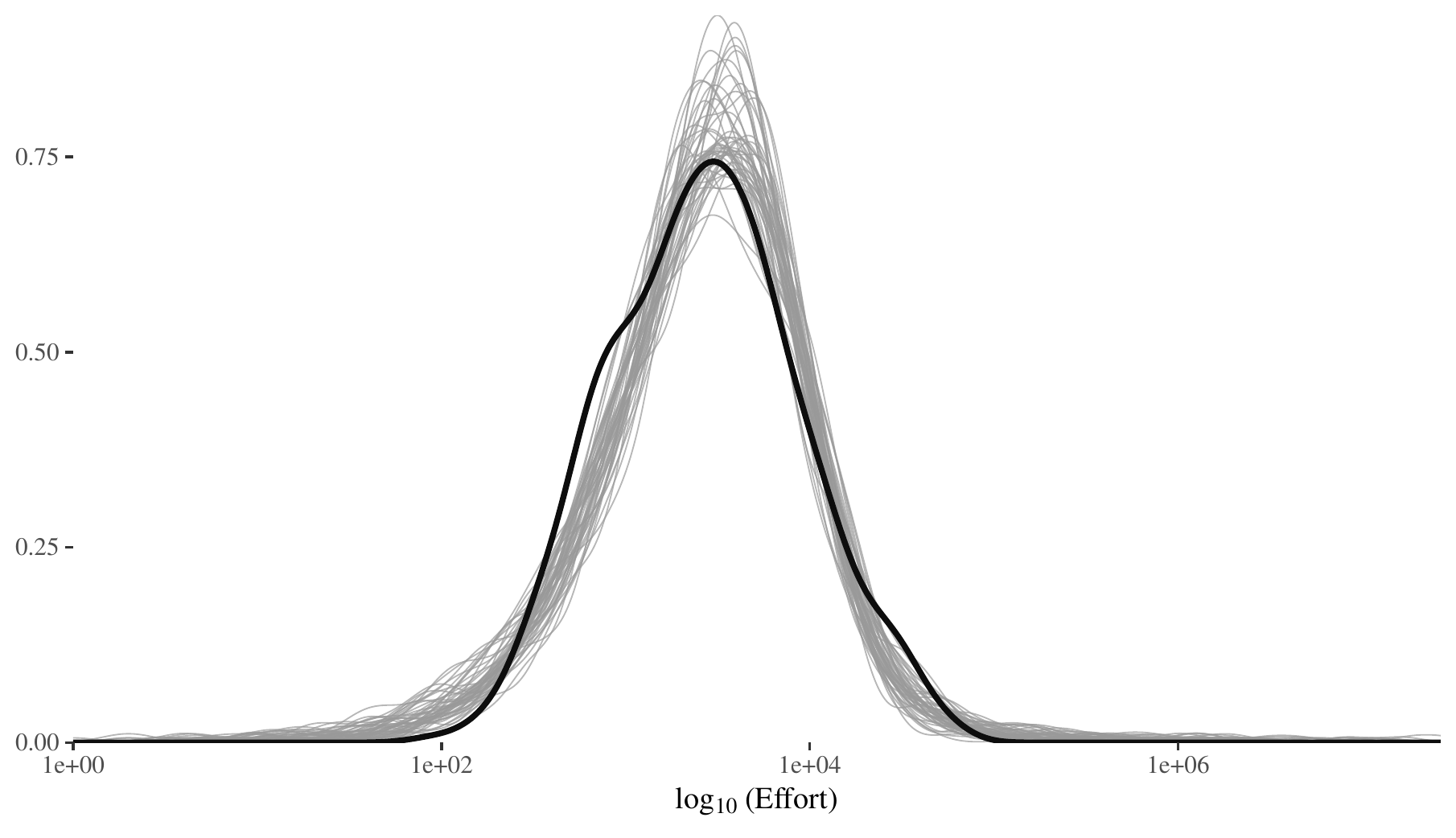}
    \caption{Comparison of the empirical distribution of the data (`AD\_clean'), to the distributions of simulated data from the posterior predictive distribution (50 samples). Note that the $x$-axis has been transformed}
    \label{fig:kdp}
\end{figure}

We sampled four chains, each with 2,000 iterations, and the first half of the iterations were discarded as warmup iterations. For our imputed data sets this means that we have $1,000 \times 4 \times 25 = 100,000$ posterior samples. Figure~\ref{fig:kdp} provides a comparison of our empirical outcome $y$ (using the data set `A\_clean'), with draws from our posterior distribution; we see evidence of a fairly good match; a perfect match is not what we want since then we could just use our data as-is, i.e., the variability of each $y_{\text{rep}}$ vis-à-vis $y$ is what interests us.

\begin{figure}
    \centering
    \includegraphics[width=\columnwidth]{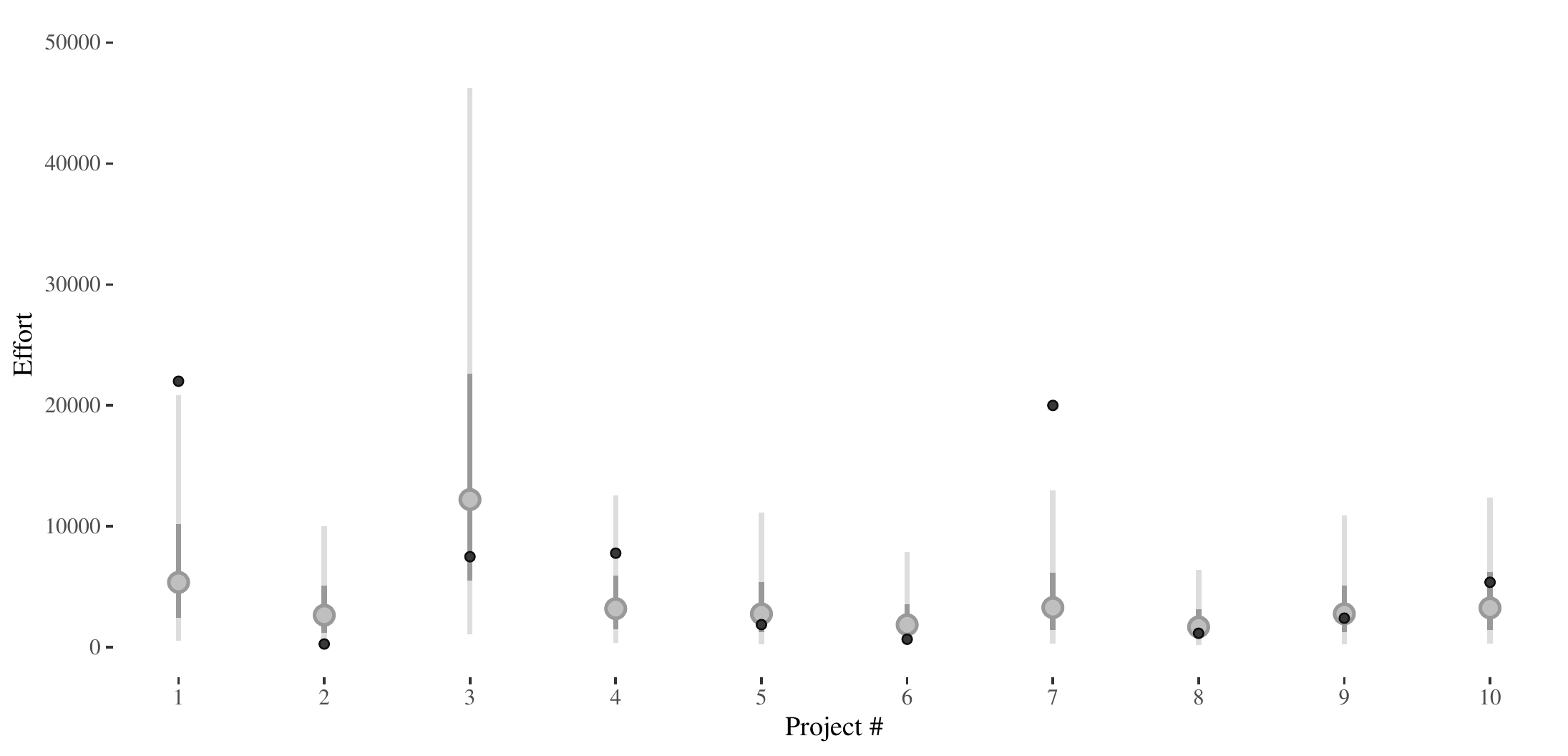}
    \caption{Posterior predictive checks of the first 10 projects in our data set. Vertical bars with points indicate the simulated medians and darker points indicate our empirical values. Thicker and thinner lines indicate 50\% and 90\% central intervals, respectively. We drew 500 samples from the posterior}
    \label{fig:ppc_case2}
\end{figure}

Another check one should do is to investigate how well the sampling ($y_{\text{rep}}$) matches our empirical data $y$ for each project (row) in our data set. In Fig.~\ref{fig:ppc_case2}, we have drawn 500 samples from the posterior. As we can see our empirical data, $y$, does not always match $y_{\text{rep}}$, but that is all fine actually, what we want is a model that \textit{on average} makes better predictions. After all, if we would want perfect predictions for our data set, why not use the data set as-is?

\subsection{Conduct Inference}\label{sec:results}
\begin{figure}
    \sidecaption
    \includegraphics[scale=0.32]{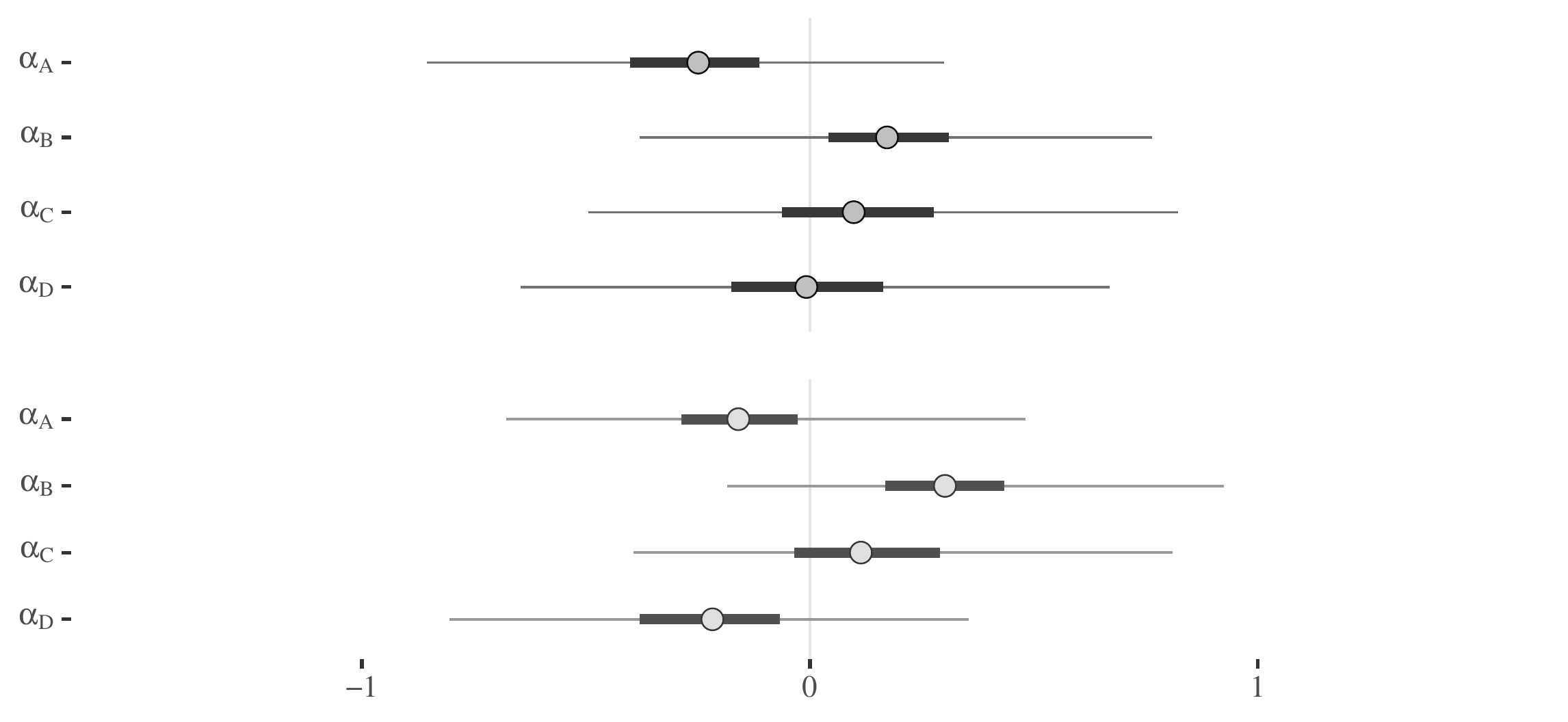}
    \caption{Interval plots of $\alpha$ estimates with 50\% and 95\% uncertainty intervals. Upper plot is the imputed data set, below plot the cleaned data set. The uncertainty is slightly different between each category even though this is not very obvious in this plot (the cleaned data set, below, has higher uncertainty)}
    \label{fig:alphas}
\end{figure}

If we turn our attention to the estimated intercepts for our group-level variable \texttt{DQR} (i.e., a project's rating according to the quality of data collected), we see something interesting in Fig.~\ref{fig:alphas}. There is a clear pattern, in both data sets, where quality rating `A' and `D' are to the left, while `B' and `C' are to the right. It is an indication that these two groups are perceived as similar to each other, which is a bit ironic since `A' and `D' are conceptually the opposite of each other in terms of data quality rating. This indicates that one should question the data quality ratings in the data set and, given enough data, \texttt{DQR} seems to become less critical.

\begin{figure}[t]
    \sidecaption[t]
    \includegraphics[scale=0.42]{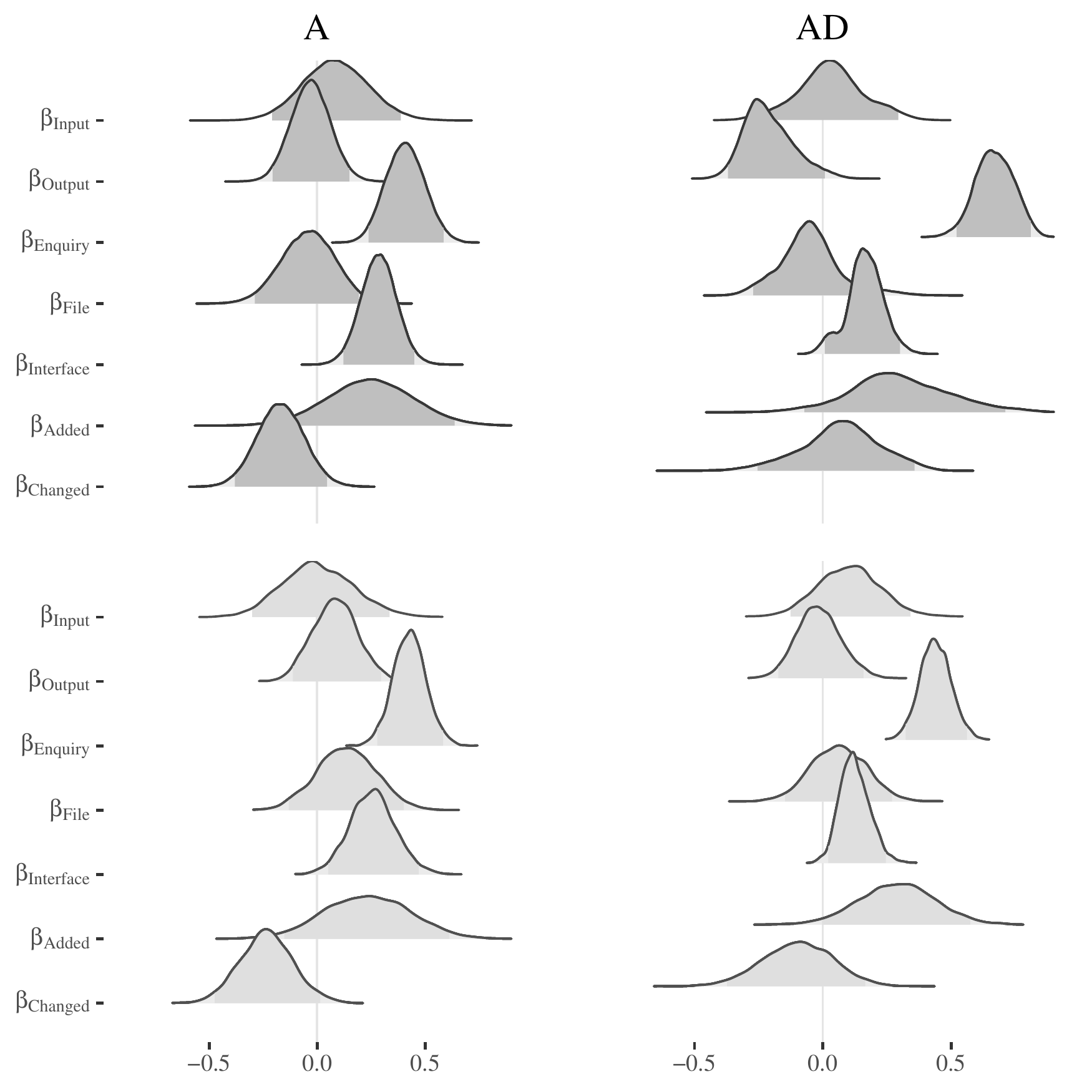}
    \caption{Density plots drawn with overlapping ridgelines of $\beta$ estimates and 95\% uncertainty intervals. Left column presents data sets `A', while right column presents data sets `AD'. Dark gray plots indicate imputed data sets (first row), while light gray represents cleaned sets. In particular the top right and lower left plots are of interest to us since they represent the data-greed approach and the state of practice, respectively}
    \label{fig:betas}
\end{figure}

Examining the estimated parameters (Fig.~\ref{fig:betas}) we see that parameters perceived as `significant' differ between the data sets.\footnote{Our notion of `significant' is here that the 95\% highest density posterior interval does not cross zero. However, other notions do exist~\citep{kruschke18rope}.} There are three comparisons we should make here. First, comparing imputed with cleaned data sets (within each column). Second, comparing `A' and `AD' models (between columns). Third, compare the upper right and lower left plots (data-greed approach vs.\ state of practice).

First, if we look at the left column, we see that nothing significant has changed. On the other hand, examining the right column, we see that $\beta_{\text{Added}}$ is no longer significant in the imputed data set. This might make you sad. However, using all data can lead to weaker inferences and, honestly, should we not make use of all data available no matter our wishful thinking concerning inferences? 

Second, if we compare our simple models with our multi-level models (between column comparisons), examining the two lower plots we see that $\beta_{\text{Added}}$ has become `significant' in the model where we make use of all quality ratings (right plot). In this particular case, one would lean towards the multi-level model (right plot) since it, after all, makes use of more data and employs partial pooling to avoid over-fitting.

Finally, we should compare the upper right and lower left plots (our data-greed approach with state of practice); they are the reason for conducting this study. Two things are worth noticing here: $(i)$ $\beta_{\text{Enquiry}}$ is shifted noticeably more to the right in the imputed data set and is significant, as is $\beta_{\text{Interface}}$. $(ii)$ $\beta_{\text{Changed}}$ is clearly not significant in the imputed data set, while in the cleaned data set, it is nearly so. Once again, making use of more data can lead to weaker inferences, which is a good thing.

Let us now examine what the posterior predictive distribution provides us with concerning point estimates regarding our outcome \texttt{Effort}. The posterior distribution allows us to set predictors at different values and generate predicted outcomes with uncertainty. In our case, we would like to examine the difference between posterior predictive distributions of `A\_clean' and `AD', since `A\_clean' is based on the assumptions commonly used in literature and `AD' makes use of as much data as possible, i.e., our data-greed approach.

\begin{figure}
    \centering
    \includegraphics[width=\columnwidth]{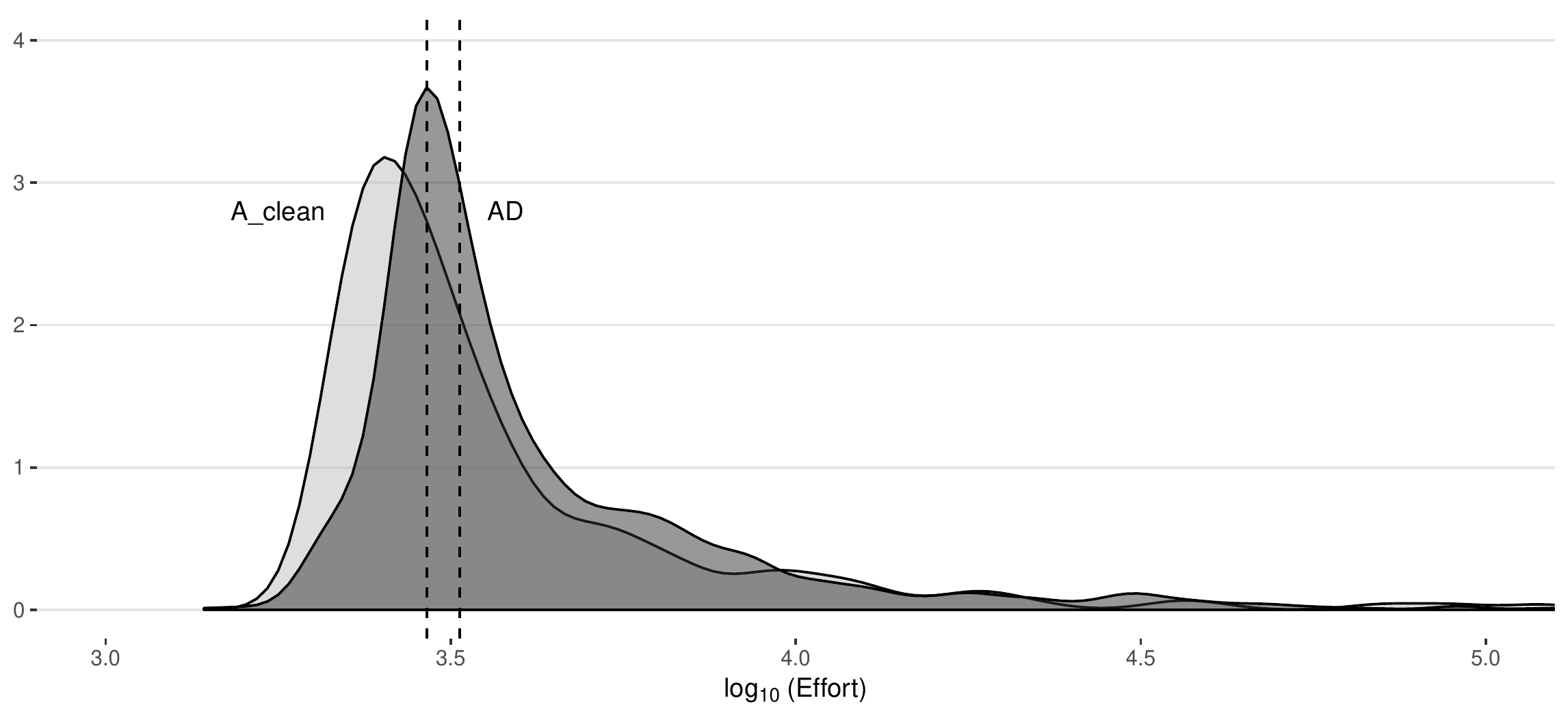}
    \caption{Comparison of the posterior distributions of the `A\_clean' and `AD' data sets ($n=4000$). Notice the transformation of the $x$-axis. The median values on natural scale, with 95\% highest posterior density intervals for `A\_clean' and `AD', are $\tilde{\mu}_{\text{A\_clean}}=2937$, $95\% \text{ HPDI}[2069,15441]$ and $\tilde{\mu}_{\text{AD}}=3280$, $95\% \text{ HPDI}[1920, 17783]$, respectively (medians indicated by vertical lines). Highest posterior density interval (HPDI) is the tightest interval containing the specified probability mass, i.e., 95\% in our case}
    \label{fig:comp_dens}
\end{figure}

Plotting our two posterior distributions indicates the differences (see Fig.~\ref{fig:comp_dens}). As is clear, the median is higher when taking into account more data (`AD'), but the uncertainty is also slightly larger. Ultimately, comparing these types of point estimates should always go hand in hand with the purpose of the analysis, i.e., the utility function.

The posterior distribution allows us to make probabilistic statements that provide us with a deeper understanding of the phenomena we study. We could make statements for each model separately, i.e., investigating `AD' we can say that the probability that the effect is greater than 10\% (a decline of \textgreater 10\%) for the \texttt{D} category is 2.6\%, while for the \texttt{A} category it is 7.2\%. Alternatively, in the case of `A\_clean', that the probability that the effect of \texttt{Enquiry} is greater than 5\% (an increase of \textgreater 5\% in this case) is 58.6\%. Of course, one could also look at the probability that the estimates of parameter $\beta_{\text{Input}}$ is larger than 0 using the `AD\_clean' and `AD' data sets, i.e., 79.8\% and 60.2\%, respectively.

These types of probability statements are a positive aspect of Bayesian analysis and the posterior probability distributions that accompanies it.
 
\subsection{Threats to Validity}\label{sec:ttv}
In this section, we will cover threats to validity from a quantitative and statistical perspective. The below threats are not the type of threats that are normally discussed in \ESE{}~\citep{wohlin12exp} (such as threats to internal and external validity). The latter refers to a rigid experimental design (often based on statistical hypothesis testing) and are mainly qualitative; in contrast, the threats we discuss are grounded in the quantitative analysis we performed, and as such they address the very design of the analysis (which allows for much more flexibility using the tools of Bayesian data analysis). In Sect.~\ref{subsubsec:commonthreats} we will compare our study's design to recent guidelines concerning the design and reporting of software analytics studies.

\textbf{Directed Acyclic Graphs (DAGs).} Making ones scientific model explicit is dangerous since it becomes open to attack. We believe, however, that it should be compulsory in any scientific publication. We employed DAGs to this end, a concept refined by Pearl and others~\citep{pearl09causality}. Using DAGs make things explicit. If things are explicit, they can be criticized. One threat to validity is of course that our scientific model is wrong and \texttt{AFP} is not a mediator (Sect.~\ref{subsec:CA}). This is for the reader to comment on. Of course, instead of using the graphical approach of DAGs, and applying $do$-calculus to determine $d$-separation, one could walk down the road of numerical approaches according to \citet{schoelkopf17ci}.

\textbf{Non-Identifiability.} Through our non-identifiability analysis we concluded that the variable \texttt{Deleted} should be removed. Removing this variable is a trade-off. The non-identifiability analysis showed that it should be removed, thus allowing for better understandability and better out of sample prediction. However, we could have taken a more prudent approach and investigated \texttt{Deleted}'s role in predictions, but in this particular case, we believe the initial analysis provided us with a convincing argument to remove it.

\textbf{Priors.} The sensitivity analysis of priors provided us with confidence regarding the choice of priors. We conducted prior predictive analysis and, together with recommendations regarding default priors, concluded that our selection of priors was balanced. However, our conclusions could be wrong, and further studies could indicate that our priors are too broad. The latter is, however, what one can expect when doing science.

\textbf{Bayesian Data Analysis.} In a Bayesian context, model comparison is often divided into three categories $\mathcal{M}$-open, $\mathcal{M}$-complete, and $\mathcal{M}$-closed~\citep{yaoVSG2018m-open, navarro2019m-open}. In the $\mathcal{M}$-open world the relationship between the data generating process and our list of models $\mathcal{M} = M_1, \ldots, M_K$ is unknown. However, what we do know is that $M_t$, our `true' model, is not in $\mathcal{M}$, and we cannot specify the explicit form $p(\tilde{y}|y)$ due to computational or conceptual reasons. In the $\mathcal{M}$-complete world, $M_t$ is also not in $\mathcal{M}$, but we use any model in $\mathcal{M}$ that is closest in Kullback-Leibler divergence~\citep{yaoVSG2018m-open,betancourt15kullback-leibler}. Finally, in the case of the $\mathcal{M}$-closed world, $M_t \in \mathcal{M}$. 

The bulk of statistical methodology is concerned with the latter category ($\mathcal{M}$-closed) and~\citet{clarkeCY2013m-open} claims that,

\begin{quotation}
this class of problems is comparatively simple and well studied.
\end{quotation}

Many problems we face are in the $\mathcal{M}$-complete and not the $\mathcal{M}$-closed world (this chapter is such an example). Selecting the `best' model is often done through relative comparisons of $M_1, \ldots, M_K$ using the Watanabe-Akaike information criterion (WAIC) or leave one out cross-validation with Pareto-smoothed importance sampling~\citep{loo}. However, to use  WAIC or PSIS-LOO for out of sample prediction, one should use the same data set for each model (e.g., you can change the likelihood and priors, but the data set is fixed). 

In this chapter, we have not done model comparison (e.g., using PSIS-LOO~\citep{loo}), but the reason is apparent---we use different data sets---which is the purpose of this chapter. To this end we defend our choice of likelihood, i.e., the Gamma-Poisson (a.k.a.\ negative binomial), epistemologically: If we have counts from zero to infinity, where the variance is significantly different from the mean then, from a maximum entropy point of view, Gamma-Poisson is a rational choice. By conducting posterior predictive checks, we ultimately received yet another validation to strengthen us in our opinion that the model has predictive capabilities (Sect.~\ref{sec:results}).

\textbf{Residuals.} One threat to validity is also the residuals of the model (fitting deviation). If the residuals are too large, it is an indication that the model does not fit data well. This is a trade-off since a perfect fit could imply overfitting. By conducting posterior predictive checks, we concluded that the models, as such, had a convincing fit (see, e.g., Fig.~\ref{fig:kdp}). However, investigating the residuals for each estimated parameter provides us with a better view.

\begin{figure}
    \sidecaption
    \includegraphics[scale=0.32]{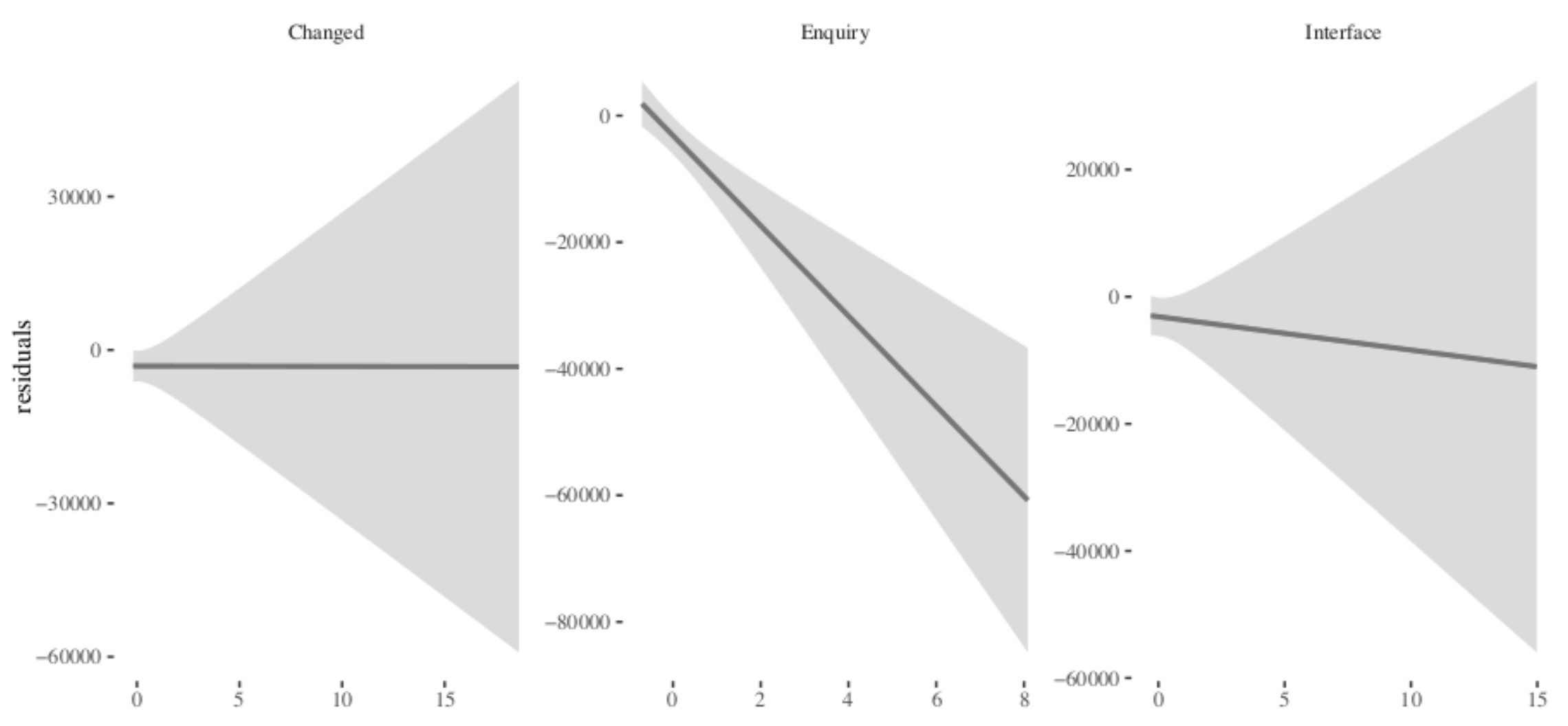}
    \caption{Residuals of three parameters of interest}
    \label{fig:resid}
\end{figure}

In Fig.~\ref{fig:resid}, we see estimates of the parameters with the most unobservable statistical error. The residuals are linear, which indicates predictability in our imputation. But, remember, \texttt{Changed} was judged to be the easiest to impute, but still \texttt{Enquiry} and \texttt{Interface} provide less uncertainty considering residuals (Sect.~\ref{subsec:MDA})---that is due to the imputation most likely. Nevertheless, in the end, we have employed multi-level models when possible and, thus, taken a conservative approach by using partial pooling. The latter should encourage us to put some trust in the model's ability to avoid overfitting, i.e., learn too much from our data.

\subsubsection{Common Threats in Software Analytics Papers} \label{subsubsec:commonthreats}
Finally, we believe that using the traditional threats to validity nomenclature seen in \ESE{} research most likely does not fit the type of studies we present here. Instead we will propose something different.

\citet{krishnaMMS2018smells} presents 12 `bad smells' in software analytics papers. Below we will now cover each `smell' and contrast it with what we did in this chapter.

\begin{enumerate}
\item \textbf{Not interesting.} (Research that has negligible software engineering impact.) We argue that the problem we have analyzed in this section is not only relevant, common, and interesting. To this we mainly point to Sect.~\ref{sec:reading}.

\item \textbf{Not using related work.} (Unawareness of related work concerning RQs and SOA.) We point to further reading (Sect.~\ref{sec:reading}) as a basis for this study, i.e., studies which throw away data, and show how other state of the art analytical approaches might be more suitable.

\item \textbf{Using deprecated and suspect data.} (Using data out of     convenience.) The data is definitely suspect as we have shown by our analysis of data quality ratings, but it would be hard to argue that the data is deprecated. However, we used a particular version of the data set, but that was due to our intention to align with previous work.

\item \textbf{Inadequate reporting.} (Partial reporting, e.g., only means.) In this section, we have presented not only point estimates but also contrasted different distributions and derived probabilistic statements. We would have also liked to provide model comparisons but, alas, the design of this study did not allow this. To this end, we rely on posterior predictive checks.

\item \textbf{Under-powered experiments.} (Small effect sizes and little theory.) Using more data provides us with more statistical power, and we base our priors on state of the art recommendations and logical conclusions, e.g., estimating that the world's population is part of a project is not appropriate.

\item \textbf{$\boldsymbol{p< 0.05}$ and all that.} (Abuse of null hypothesis testing.) We mention $p$-values only when making a point not to use them.

\item \textbf{Assumptions of normality and equal variances.} (impact of outliers and heteroscedasticity.) We use a Bayesian generalized linear model, which we model using a Gamma-Poisson likelihood. Additionally, we employ multi-level models when possible, and hence make use of partial pooling (which takes into account the presence of outliers).

\item \textbf{Not exploring stability.} We conducted a sensitivity analysis of priors, and we report on the differences between imputed and cleaned data sets.

\item \textbf{No data visualization.} We leave it up to the reader to decide if appropriate levels of visualization were used. We have followed guidelines on data visualization~\citep{GabrySVBG17}.

\item \textbf{Not tuning.} We avoid bias in comparisons mainly by clearly stating our assumptions, conducting a sensitivity analysis, making use of multi-level models, and, generally speaking, following guidelines on how to conduct Bayesian data analysis.

\item \textbf{Not exploring simplicity.} Using state of the art missing data analysis is needed and wanted to decrease our bias. Additionally, using a complex mixture model was unavoidable because of epistemological reasons, as presented earlier in this section. We used simulated data to assess the appropriateness of our likelihood and priors independently.

\item \textbf{Not justifying choice of learner.} This concerns, ultimately, the risk of over-estimation (or over-fitting). We would argue that any usage of frequentist statistics would potentially introduce this `smell', i.e., using uniform priors, as is the case in a traditional frequentist setting, ensures maximum over-fitting.
\end{enumerate}

\subsection{Discussion}\label{sec:disc}
We argue that one should have solid reasons to throw away data since we now have techniques available that can provide us with the opportunity to use as much data as possible. The example we provided showed that by using missing data techniques, in combination with Bayesian multi-level models, we could better make use of the data and, thus, gain higher confidence concerning our findings. The inferences can become weaker, but ask yourself if that is not how you think your fellow researchers should conduct their analysis. We could show two things in our analysis: $(i)$ Parameters' `significance' changed depending on if we used imputation or not, and $(ii)$ there was really not much of a difference between the various data quality ratings (once again indicating that we should use as much data as possible).

However, we pay a price for this more sophisticated analysis. It is a more involved analysis compared to a frequentist analysis where only the likelihood of the outcome is required to be specified, and maximum likelihood estimates are not conditioned on the observed outcome; the uncertainty is instead connected to the sampling distribution of the estimator. The same applies to confidence intervals in the frequentist world, i.e., one can set up a distribution of predictions, but it entails repeating the process of random sampling on which we apply the estimator every time to then generate point predictions. Contrast this with conditioning on the posterior predictive distribution, which is based on the observed outcome. 

Additionally, making probabilistic statements is very much more natural when having a posterior at hand, while the $p$-values we have made use of traditionally rely on observing a $z$-statistic that is so large (in magnitude) if the null hypothesis is true, i.e., not if the scientific hypothesis is true. To make the point, the term `$p$-value' was used in this section for the first time here in this section and in our case, where we used different data sets, one could have expected us to lean towards traditional hypothesis testing, since it was not possible to compare models explicitly, regarding out of sample predictions.

We will not further contrast our approach with how analyses are done in \ESE{} today. Suffice to say, issues such as the arbitrary $\alpha=.05$ cut-off, the usage of null hypothesis significance testing and the reliance on confidence intervals have been criticized \citep{ioannidis05false, MoreyHRLW2016CI, Nuzzo14errors, Woolston15pvalues}, and when analyzing the arguments, we have concluded that many of the issues plaguing other scientific fields are equally relevant to come to terms with in \ESE{}.

\begin{shaded}
We believe that evidence-based interpretation is more straightforward with Bayesian data analysis, and \ESE{} should embrace it as soon as possible. In our view, it is a natural choice to make in this particular case; to base one's inferences on more data is wise, and doing so in a Bayesian context is natural.
\end{shaded}

\section{Recommended Further Reading}\label{sec:reading}
There are few early publications in software engineering where we see evidence of using MLMs. In \citep{ehrlich12MLM} the authors used multilevel models for assessing communication in global software development, while in~\citep{hassan2017MLM}, the authors applied MLMs for studying reviews in an app store. However, both studies used a frequentist approach (maximum likelihood), i.e., not a Bayesian approach.

As far as we can tell, there are only a few examples of studies in software engineering that have applied BDA with MLMs to this date~\citep{furia16bayes,ernst18MLM}. \citet{furia16bayes} presents several cases of how BDA could be used in computer science and software engineering research. In particular, the aspects of including prior belief\slash knowledge in MLMs are emphasized, which is further elaborated on in~\citep{furiaFT20bayes}. \citet{ernst18MLM}, on the other hand, presents a conceptual replication of an existing study where he shows that MLMs support cross-project comparisons while preserving local context, mainly through the concept of partial pooling, as used in Bayesian MLMs.\footnote{Partial pooling takes into account variance between units.}

Finally, much literature on BDA exist, but not all have the clarity that is needed to explain, sometimes, relatively complex concepts. If one would like to read up on the basics of probability and Bayesian statistics we recommend~\citep{jaynes03prob}, for a slightly more in-depth view of Bayesian statistics we would recommend~\citep{lambert2018student}. For a hands-on approach to BDA, we recommend~\citep{mcelreath15statrethink}; McElreath's book \textit{Statistical Rethinking: A Bayesian Course with Examples in R and Stan} is an example of how seemingly complex issues can be explained beautifully, while at the same time help the reader improve their skills in BDA\@. To conclude, there is one book that every researcher should have on their shelf, \textit{Bayesian Data Analysis} by \citet{gelman2013bayesian}, which is considered the leading text on Bayesian methods.

Missing data can be handled in two main ways. Either we delete data using one of three main approaches (listwise or pairwise deletion, and column deletion) or we impute new data. Concerning missing data, we conclude the matter is not new to the \ESE{} community. \citet{liebchenS08dataquality} have pointed out that the community needs more research into ways of identifying and repairing noisy cases, and \citet{mockus2008mi} claims that the standard approach to handle missing data, i.e., remove cases of missing values (e.g., listwise deletion), is prevalent in \ESE{}.

Two additional studies on missing data are, however, worthwhile pointing out. \citet{myrtveitSO01mi} investigated listwise deletion, mean imputation, similar response pattern imputation, and full information maximum likelihood (FIML), and conclude that FIML is the only technique appropriate when data are not missing completely at random. Finally, \citet{cartwrightSS03mi} conclude that $k$-nearest neighbor and sample mean imputation are significantly better in improving model fit when dealing with imputation. However, much has happened concerning research in imputation techniques lately.

In this chapter, we focused on multivariate imputation by chained equations (MICE), sometimes called fully conditional specification or sequential regression multiple imputation, a technique that has emerged as a principled method of dealing with missing data during the last decade~\citep{vanBuuren07mice}. MICE specifies a multivariate imputation model on a variable-by-variable basis by a set of conditional densities (one for each incomplete variable) and draws imputations by reiterating the conditional densities. The original idea behind MICE is old, see, e.g., stochastic relaxation~\citep{gemanG84stochrelax}, but the recent refinements and implementations have made the technique easily accessible.

Finally, related work concerning the ISBSG data set is worthwhile pointing out. \citet{fernandez14sysrevISBSG} present pros and cons of using the ISBSG data set. A systematic mapping review was used as the research method, which was applied to over 120 papers. The dependent variable of interest was usually \texttt{Effort} (more than 70\% of the studies), and the most frequently used methods were regression ($\sim$60\%) and machine learning ($\sim$35\%), the latter a term where many techniques can hide. Worth noting is also that Release 10 was used most frequently, which provided us with a reason to also use that data set. Additionally, we also used \texttt{Effort} as the dependent variable of interest, since a majority of the studies seem to find that variable interesting to study. (The importance of the ISBSG data set, when considering replication studies in \ESE{}, has already been pointed out by \citet{shepperdAC18repl}.) By and large, our chapter took another approach entirely, we imputed missing data in a Bayesian context, and we see this more as complementary to some of the work mentioned above.

\section{Conclusion}\label{sec:concl}
In this chapter, we introduced the reader to Bayesian data analysis. Before even designing the model, we took several steps, each providing us with a better understanding of the data. We did a causal analysis, analyzed non-identifiability, performed sensitivity analysis of priors, and an analysis of missing data. Additionally, we presented the reader with several diagnostics one should use for sanity checking a statistical model. Except for the missing data analysis (if no missingness is present), we would argue that this is something one should always do when conducting Bayesian data analysis.

Missing data was an additional complexity that our case presented. We recommend that one should always be conservative with throwing away data. Many state of the art techniques exist today, which provides the researcher with ample of possibilities to conduct rigorous, systematic, and transparent missing data analysis. We followed a traditional imputation approach, but other approaches, i.e., purely Bayesian, do exist. In our example, we showed that inferences can become weaker, which is not a bad thing, and that the qualitative assessment of quality ratings can be biased. This further strengthens the argument never to throw data away.

By using Bayesian data analysis, we believe that researchers will be able to get a more nuanced view of the challenges they are investigating. In short, we do not need $p$-values for this.

%\section*{Appendix}
%\addcontentsline{toc}{section}{Appendix}

\bibliographystyle{spbasic}
\bibliography{11_torkar_BDA}
\end{document}